\documentclass[11pt]{article}

\usepackage{amsmath,amssymb}           
\usepackage{amscd}                     
\usepackage{epsfig}                    
\usepackage[matrix,arrow]{xy}          
\usepackage{xspace}                    
\usepackage{stmaryrd}                  
\usepackage{slashed}
\usepackage{tikz}
\usepackage{pbox}

\usepackage{jheppub}                   
\makeatletter
\gdef\@fpheader{\ }                    
\makeatother

\DeclareSymbolFont{bbold}{U}{bbold}{m}{n}
\DeclareSymbolFontAlphabet{\mathbbold}{bbold}

\newcommand{\bq}{\begin{equation}}
\newcommand{\eq}{\end{equation}}
\newcommand{\bea}{\begin{eqnarray}}
\newcommand{\eea}{\end{eqnarray}}

\newcommand{\dd}{\mathrm{d}}
\newcommand{\ee}{\mathrm{e}}
\newcommand{\ii}{\mathrm{i}}

\newcommand{\der}{\partial}

\newcommand{\bbZ}{\mathbb{Z}}
\newcommand{\bbR}{\mathbb{R}}

\DeclareMathOperator{\SU}{\mathit{SU}}
\DeclareMathOperator{\SO}{\mathit{SO}}
\DeclareMathOperator{\SL}{\mathit{SL}}
\DeclareMathOperator{\GL}{\mathit{GL}}
\DeclareMathOperator{\USp}{\mathit{USp}}

\DeclareMathOperator{\Spin}{\mathit{Spin}}

\DeclareMathOperator{\gl}{\mathit{gl}}

\DeclareMathOperator{\Cliff}{Cliff}

\newcommand{\rep}[1]{\mathbf{#1}}
\newcommand{\repp}[2]{(\rep{#1}, \rep{#2})}
\newcommand{\id}{\mathbbold{1}}

\DeclareMathOperator{\vol}{vol}

\newcommand{\Gs}[1]{\Gamma(#1)}

\newcommand{\ph}[1]{\phantom{#1}}

\newcommand{\Lgen}{L}

\newcommand{\Bgen}[2]{\left\llbracket#1,#2\right\rrbracket}
\newcommand{\BLie}[2]{\left[#1,#2\right]}

\newcommand{\Dgen}{{D}}

\DeclareMathOperator{\adj}{ad}

\newcommand{\LC}{\nabla}

\newcommand{\Ric}{\mathcal{R}}
\newcommand{\Scalar}{\mathcal{R}}

\newcommand{\GenRic}{R^{\scriptscriptstyle 0}}
\newcommand{\GenRicci}{R}
\newcommand{\GenS}{R}

\DeclareMathOperator{\Diff}{Diff}

\newcommand{\proj}[1]{\times_{#1}}

\newcommand{\inn}{\mathbin{\lrcorner}}
\newcommand{\oadj}{\proj{\text{ad}}}
\newcommand{\on}{\proj{N}}

\newcommand{\Ggeom}{G_{\textrm{split}}}

\DeclareMathOperator{\Edd}{\mathit{E_{d(d)}}}
\DeclareMathOperator{\Hd}{\mathit{H_d}}
\DeclareMathOperator{\dHd}{\mathit{\tilde{H}_d}}
\DeclareMathOperator{\E7}{\mathit{E_{7(7)}}}

\newcommand{\Lodd}{\Lambda^\text{odd}}
\newcommand{\Leven}{\Lambda^\text{even}}
\newcommand{\Leo}{\Lambda^\text{even/odd}}

\newcommand{\tA}{{\tilde{A}}}
\newcommand{\tF}{{\tilde{F}}}
\newcommand{\ta}{{\tilde{a}}}
\newcommand{\talpha}{{\tilde{\alpha}}}
\newcommand{\tLambda}{{\tilde{\Lambda}}}

\newcommand{\am}{Q}
\newcommand{\dex}{\hat{c}}

\title{$\Edd\times\bbR^+$ Generalised Geometry, Connections \\
   and M theory}

\author{Andr\'e Coimbra, 
  Charles Strickland-Constable  
  and Daniel Waldram}
\affiliation{Department of Physics,
   Imperial College London \\
   Prince Consort Road, London, SW7 2AZ, UK}

\emailAdd{a.coimbra08@imperial.ac.uk}
\emailAdd{charles.strickland-constable08@imperial.ac.uk}
\emailAdd{d.waldram@imperial.ac.uk}

\subheader{\textrm{Imperial/TP/11/DW/02}}

\abstract{
We show that generalised geometry gives a unified
description of bosonic eleven-dimensional supergravity restricted to a
$d$-dimensional manifold for all $d\leq7$. The theory is based on an  
extended tangent space which admits a natural $\Edd\times\bbR^+$
action. The bosonic degrees of freedom are unified as a ``generalised
metric'', as are the diffeomorphism and gauge symmetries, while the
local $O(d)$ symmetry is promoted to $\Hd$, the maximally compact
subgroup of $\Edd$. We introduce the analogue of the Levi--Civita
connection and the Ricci tensor and show that the bosonic action and
equations of motion are simply given by the generalised Ricci scalar
and the vanishing of the generalised Ricci tensor respectively. The
formalism also gives a unified description of the bosonic NSNS and RR
sectors of type II supergravity in $d-1$ dimensions. Locally the
formulation also describes M-theory variants of double field theory
and we derive the corresponding section condition in general
dimension. We comment on the relation to other approaches to M theory
with $\Edd$ symmetry, as well as the connections to flux
compactifications and the embedding tensor formalism.}

\begin{document}
\maketitle


\section{Introduction}
\label{sec:intro}

The idea that eleven-dimensional supergravity, or for that matter M
theory, might have a more unified description incorporating a larger
symmetry group is a long-standing one. Following the original
observation that the dimensionally reduced supergravity has a hidden
$\Edd$ global symmetry~\cite{CJ,cremmer,julia}, formulations using
exceptional groups, as well as their infinite-dimensional extensions,
have appeared in various
guises~\cite{deWN,duff,Nic,KNS,deWN-extra,West-conj,e10,hillmann,BP4d,%
BP5d,thompson,BP-alg,BPW}. 

In this paper we show that generalised geometry~\cite{GCY,Gualtieri}
gives a unified geometrical description of bosonic eleven-dimensional
supergravity restricted to a $d$-dimensional manifold for
$d\leq7$. One starts with an extended tangent
space~\cite{chris,PW} which admits a natural $\Edd\times\bbR^+$
action. The bosonic degrees of freedom are unified as a ``generalised
metric'' $G$, while the diffeomorphism and gauge symmetries are
encoded as a ``generalised Lie derivative." The local $O(d)$ symmetry
is promoted to $\Hd$, the maximally compact subgroup of
$\Edd$. Remarkably, the dynamics are simply the generalised
geometrical analogue of Einstein gravity. The bosonic action is given
by  
\begin{equation}
\label{eq:S-intro}
   S_{\text{B}} = \int \vol_G \GenS , 
\end{equation}
where $\vol_G$ is the volume form associated to the generalised metric 
and $\GenS$ is the analogue of the Ricci scalar. The 
corresponding equations of motion are simply
\begin{equation}
\label{eq:eom-intro}
   \GenRicci_{MN} = 0 ,
\end{equation}
where $\GenRicci_{MN}$ is the analogue of the Ricci tensor. This work extends
the corresponding description of type II theories in terms of
$O(10,10)\times\bbR^+$ generalised geometry given in~\cite{CSW1}. 

The formalism also describes type II theories
restricted to $d-1$ dimensions, geometrising not only the NSNS sector
but also the RR fields. Even though here we focus our attention on the
bosonic sector, we will find that, in fact, the supersymmetry
variations of the fermions are already encoded by the geometry. In a
forthcoming paper~\cite{CSW2} we extend the construction 
to include the fermion fields to leading order, thus completing the
reformulation of restricted eleven-dimensional supergravity.

That eleven-dimensional supergravity could be reformulated with a manifest
local $H_7=\SU(8)/\bbZ_2$ symmetry, and fields transforming in
$E_{7(7)}$ representations was first shown by de Wit and
Nicolai~\cite{deWN}, who also conjectured that
formulations using other $\Edd$ groups should exist. This was 
elaborated on in~\cite{Nic,KNS} for the case of $H_8=SO(16)$ local
symmetry and $E_{8(8)}$ representations.  Julia~\cite{julia} had
earlier noted that for dimensional reductions to three-dimensions the
global $E_{8(8)}$ symmetry includes part of the three-dimensional
helicity group and wondered if $E_{8(8)}$ could be a symmetry of the
theory in all dimensions, while Duff~\cite{duff} also independently
conjectured that $E_{8(8)}$ was a global symmetry of the
eleven-dimensional equations of motion. The construction of
$\Edd\times\bbR^+$ generalised geometry can be viewed as providing a
geometrical basis for these results for $d\leq7$. (Note that the
relevant $\Edd$ action is by the continuous group rather than the
discrete U-duality group that appears in toroidal reductions of
M-theory~\cite{HT}.) 

The idea that dimensional reductions to less than three dimensions
should realise infinite dimensional Kac--Moody algebras was first
proposed in~\cite{julia,julia-KM}. The case of $E_9$ was analysed
in~\cite{N-E9,NW}, while $E_{10}$ was discussed
in~\cite{N-hyperbolic,Miz,DH}. That such algebras might appear as
symmetries of or in classifying the degrees of freedom of
the uncompactified theory is mentioned
in~\cite{julia,deWN,duff}. It is West~\cite{West-conj} who
was the first to conjecture that $E_{11}$ is a symmetry of the
full eleven-dimensional theory, and to give a proposal for how it is
realised. At around the same time, Damour, Henneaux and
Nicolai~\cite{e10} introduced an $E_{10}$ description of the full
theory, showing that there is a coset formulation of the small-tension
expansion near a spacelike singularity.  

In West's $E_{11}$ proposal~\cite{West-conj}, the symmetry is realised
non-linearly over an extended spacetime with an infinite number of
coordinates~\cite{west-coord}. The corresponding $E_{7(7)}$ non-linear
realisation, following the construction of West and using the finite
extended spacetime originally conjectured in~\cite{deWN-extra}, has
been discussed in considerable detail by
Hillmann~\cite{hillmann}. Truncating to conventional spacetime, he was
able to show an equivalence with~\cite{deWN}, and, again, the current
paper can be viewed as the corresponding geometrical formulation,
analogous to  the relation between gravity as Riemannian geometry and
as a non-linear realisation of $\GL(4)\ltimes\bbR^4$ introduced by
Borisov and Ogievetsky~\cite{BO}.  

A related approach to realising $\Edd$ symmetries is based on
the double field theory of Hull and Zwiebach~\cite{dft}, which
describes string backgrounds in terms of fields on a doubled
spacetime that admits an action of $O(d,d)$, and also connects to 
earlier work by Duff~\cite{duff-string}, Tseytlin~\cite{tseytlin} and
Siegel~\cite{siegel}.  The dynamics~\cite{dft2} are ultimately encoded
in a version of a curvature tensor (first constructed by
Siegel~\cite{siegel} and introduced from a different perspective
in~\cite{JLP}) provided the fields are required to satisfy the
``strong constraint'', or ``section condition''. This implies that 
they depend on only half the coordinates, so locally the theory is
equivalent to the $O(d,d)\times\bbR^+$ generalised geometry
described in~\cite{CSW1}. (Interestingly, in the double field theory
realisations of the mass-deformed type IIA theory~\cite{dft-mass} and
of generic Scherk--Schwarz reductions~\cite{dft-ss} the strong
constraint can be slightly weakened, and so the relation to
generalised geometry becomes less clear.) The corresponding
formulation of M theory with $\Edd$ groups was introduced by Berman
and Perry~\cite{BP4d} for the case of $d=4$, following earlier work by
Duff and Lu~\cite{duff-lu} (see also~\cite{OL}). This was extended to
$d=5$ in~\cite{BP5d} and subsequently to $d=6,7$ in~\cite{BPW}, using
the $E_{11}$ non-linear formalism of~\cite{West-conj} (while the
relation to $O(d,d)$ double field theory was discussed
in~\cite{thompson}). In these papers a bosonic action is constructed in
terms of first-order derivatives of the generalised metric in a
generic $\Edd$ form by brute force. Arbitrary coefficients are fixed
by requiring diffeomorphism invariance upon restriction to
dependence on $d$ coordinates, and the resulting expression matches
the supergravity action up to integration by parts. This coordinate
restriction means that locally the generalised geometrical theory
constructed here is equally applicable to the double field theory
approach to M theory. In this work we are able to derive the $\Edd$
form of the action directly in terms of the scalar curvature of the
generalised connection, which is therefore automatically
invariant. Furthermore, we find a generic $\Edd$ covariant form of the
``section condition''~\cite{BP-alg} that encodes the restriction of the M
theory version of double field theory to $d$ coordinates. 

At their core, generalised geometries\footnote{Note that the term
   ``generalised geometry'' is sometimes used to refer to formulations
   where spacetime is extended to include more coordinates. Although
   the two notions are closely related, here we will limit it to the
   narrow sense of structures on a Courant algebroid as first
   introduced by Hitchin and Gualtieri~\cite{GCY,Gualtieri}, and the
   related extensions relevant to M theory.}~\cite{GCY,Gualtieri} rely
on the idea of extending the tangent space of a manifold $M$, such
that it can accommodate a larger symmetry group that includes not only
diffeomorphisms but also the gauge transformations of
supergravity. In its original form, one studies structures
on a generalised tangent space $E\simeq TM\oplus T^*M$, with a
symmetry group combining diffeomorphisms with the gauge transformations
of a two-form potential $B$. There is a natural $O(d,d)$
structure on $E$, where $d$ is the dimension of $M$, and a natural
bracket between generalised vectors giving $E$ the structure of a
Courant algebroid~\cite{roytenberg}. Slightly extending the structure
group to $O(d,d)\times\bbR^+$, we showed in an earlier
paper~\cite{CSW1} that generalised geometry gives a natural rewriting
of type II supergravity unifying the NSNS fields as a generalised
metric preserved by an $O(9,1)\times O(1,9)$ subgroup, which then
becomes a manifest local symmetry of the theory.  

The original version of generalised geometry was extended by
Hull~\cite{chris} and Pacheco and Waldram~\cite{PW} to include the
symmetries appearing in M theory. This gives a generalised tangent 
space $E  \simeq TM \oplus \Lambda^2T^*M \oplus \Lambda^5T^*M \oplus
(T^*M \otimes\Lambda^7T^*M)$, relevant to eleven-dimensional
supergravity restricted to $d\leq 7$ dimensions and admitting a
natural $E_{d(d)}$ structure. One can construct the corresponding
generalised metric and also the analogue of the Courant
bracket. Applied to type II theories, it allowed the geometrisation of
the RR fields and was then used to study the origin of general
gaugings of supergravity~\cite{E7-flux} and to reformulate the
effective theory of generic supergravity compactifications to four
dimensions as well as the conditions for existence of a supersymmetric
background~\cite{GLSW,GO}.  

The generalised tangent space contains objects
familiar from Riemannian geometry, namely a bracket structure, covariant
derivatives, torsion, and by introducing the generalised metric, the
analogue of the Levi--Civita connection, and curvature tensors. Still,
there are important differences with respect to ordinary geometry,
such as the failure of the generalised bracket to satisfy the Jacobi
identity and the fact that, unlike the Levi--Civita connection, there is a
family of torsion-free, metric-compatible generalised connections. We
discuss all these concepts for $\Edd\times\bbR^+$ generalised geometry
with a local compact subgroup $\Hd \subset \Edd$ in a manner that
treats all dimensions uniformly by decomposing under the appropriate
$\GL(d,\bbR)$ and $O(d)$ subgroups. By constructing the natural
generalised geometrical equivalent of Einstein gravity, we then find
that it contains the entire bosonic supergravity field content --
metric, warp factor, three- and dual six-form gauge fields -- and
precisely describes eleven-dimensional supergravity reduced to $d$
dimensions in the simple forms~\eqref{eq:S-intro}
and~\eqref{eq:eom-intro}. 

The paper is arranged as follows. In section~\ref{sec:EddGG} we describe the key concepts of
$\Edd\times\bbR^+$ generalised geometry, including the generalised
tangent bundle, its differential structure and the notions of
generalised connection and torsion. Next, in section~\ref{sec:Hd} we introduce
the local $\Hd$ structure and show that one can always construct a
torsion-free, $\Hd$-compatible generalised connection 
$\Dgen$, the analogue of the Levi--Civita connection. Finally, in
section~\ref{sec:sugra-rel} we review the bosonic sector of
restrictions of eleven-dimensional supergravity and show that it can
be reformulated in terms of the generalised geometry. We also comment
on the relation to type II theories, generic flux compactifications
and the embedding tensor formalism of gauged
supergravity~\cite{embedT1,embedT2}. We conclude with some summary and
discussion in section~\ref{sec:conc}.  


\section{$\Edd\times\bbR^+$ generalised geometry}
\label{sec:EddGG}

 
Following closely the construction given in section \textbf{3}
of~\cite{CSW1}, we introduce the generalised geometry versions of
the tangent space, frame bundle, Lie derivative, connections and
torsion, now in the more subtle context of an $\Edd\times\bbR^+$
structure. The $\Edd$ generalised tangent space was first developed
in~\cite{chris} and independently in~\cite{PW}, where the exceptional
Courant bracket was also given for the first time. We slightly
generalise those notions by introducing an $\bbR^+$ factor, known as
the ``trombone symmetry''~\cite{trombone}, as it allows one to specify
the isomorphism between the generalised tangent space and a sum of
vectors and forms. Physically, it is known to be related to the ``warp
factor'' of warped supergravity reductions. The need for this extra factor in the context of
$E_{7(7)}$ geometries has already been identified in~\cite{BPW,hillmann,baraglia}. 


\subsection{Generalised bundles and frames}
\label{sec:gen-bundles}


\subsubsection{Generalised tangent space}
\label{sec:gen-tangent}

We start by recalling the definition of the generalised tangent space for $\Edd\times\bbR^+$ generalised geometry~\cite{chris,PW} and defining
what is meant by the ``generalised structure''.

Let $M$ be a $d$-dimensional spin manifold with $d\leq 7$. The
generalised tangent space is isomorphic to a sum of tensor bundles
\begin{equation}
\label{eq:Eiso}
   E \simeq TM \oplus \Lambda^2T^*M \oplus \Lambda^5T^*M
         \oplus (T^*M \otimes\Lambda^7T^*M) ,
\end{equation}
where for $d<7$ some of these terms will of course be absent. The
isomorphism is not unique. The bundle is actually described using a
specific patching. If we write
\begin{equation}
\begin{aligned}
   V_{(i)} &= v_{(i)} + \omega_{(i)} + \sigma_{(i)} + \tau_{(i)}  \\
      &\in \Gs{TU_i \oplus \Lambda^2T^*U_i \oplus \Lambda^5T^*U_i 
            \oplus (T^*U_i \otimes\Lambda^7T^*U_i)} ,
\end{aligned}
\end{equation}
for a section of $E$ over the patch $U_i$, then 
\begin{equation}
\label{eq:Epatch}
   V_{(i)} = \ee^{\dd\Lambda_{(ij)}+\dd\tLambda_{(ij)}} V_{(j)} ,
\end{equation}
on the overlap $U_i\cap U_j$ where $\Lambda_{(ij)}$ and
$\tLambda_{(ij)}$ are locally two- and five-forms respectively. The
exponentiated action is given by 
\begin{equation}
\begin{aligned}
   v_{(i)} &= v_{(j)} , \\
   \omega_{(i)} &= \omega_{(j)}
      + i_{v_{(j)}}\dd\Lambda_{(ij)} , \\
   \sigma_{(i)} &= \sigma_{(j)}
      + \dd\Lambda_{(ij)}\wedge\omega_{(j)}
      + \tfrac{1}{2}\dd\Lambda_{(ij)}\wedge i_{v_{(j)}}\dd\Lambda_{(ij)}
      + i_{v_{(j)}}\dd\tLambda_{(ij)}, \\
   \tau_{(i)} &= \tau_{(j)}
      + j\dd\Lambda_{(ij)}\wedge\sigma_{(j)}
      - j\dd\tLambda_{(ij)}\wedge\omega_{(j)}
      + j\dd\Lambda_{(ij)}\wedge i_{v_{(j)}}\dd\tLambda_{(ij)} 
      \\ &\qquad {}
      + \tfrac{1}{2}j\dd\Lambda_{(ij)}\wedge\dd\Lambda_{(ij)}
         \wedge\omega_{(j)}
      + \tfrac{1}{6}j\dd\Lambda_{(ij)}\wedge\dd\Lambda_{(ij)}
         \wedge i_{v_{(j)}}\dd\Lambda_{(ij)} , 
\end{aligned}
\end{equation}
where we are using the notation of~\eqref{eq:jdef}. Technically this
defines $E$ as a result of a series of extensions 
\begin{equation}
\label{eq:twistE}
\begin{aligned}
   0 \longrightarrow \Lambda^2T^*M \longrightarrow E'' &
      \longrightarrow TM \longrightarrow 0 , \\
   0 \longrightarrow \Lambda^5T^*M \longrightarrow E' & 
      \longrightarrow E'' \longrightarrow 0 , \\
   0 \longrightarrow T^*M\otimes \Lambda^7T^*M
      \longrightarrow E &
      \longrightarrow E' \longrightarrow 0 .
\end{aligned}
\end{equation}
Note that while the $v_{(i)}$ globally are equivalent to a choice of
vector, the $\omega_{(i)}$, $\sigma_{(i)}$ and $\tau_{(i)}$ are not
globally tensors. 

Note that the collection $\Lambda_{(ij)}$
formally define a ``connective structures on gerbe'' (for a review see, for
example, \cite{gerbes}). This essentially means there is a hierarchy of
successive gauge transformations on the multiple intersections
\begin{equation}
\begin{aligned}
   \Lambda_{(ij)} + \Lambda_{(jk)}
      + \Lambda_{(ki)} &= \dd\Lambda_{(ijk)}
      &&
      \quad \text{on $U_i\cap U_j \cap U_k$}, \\
   \Lambda_{(jkl)} - \Lambda_{(ikl)}
      + \Lambda_{(ijl)} - \Lambda_{(ijk)}
      &= \dd\Lambda_{(ijkl)}
      &&
      \quad \text{on $U_i\cap U_j \cap U_k
         \cap U_l$} .
\end{aligned}
\end{equation}
If the supergravity flux is quantised, we will have
$g_{(ijkl)}=\ee^{\ii\Lambda_{(ijkl)}}\in U(1)$ with the cocycle condition
\begin{equation}
   g_{(jklm)}g^{-1}_{(iklm)}
      g_{(ijlm)}g^{-1}_{(ijkm)}
      g_{(ijkl)} = 1 ,
\end{equation}
on $U_i\cap \dots \cap U_{m}$. For $\tLambda_{(ij)}$ there is
a similar set of structures,
\begin{equation}
\begin{aligned}
   \tLambda_{(ij)} - &\tLambda_{(ik)} + \tLambda_{(jk)} \\
   	&= \dd\tLambda_{(ijk)} + \tfrac12 \tfrac{1}{3!} \Big( \Lambda_{(ij)}\wedge\dd\Lambda_{(jk)}
	+ \text{antisymmetrisation in } [ijk] \Big) \\
	& \qquad \text{on $U_i\cap U_j \cap U_k$}, \\
   \tLambda_{(ijk)} - &\tLambda_{(ijl)} + \tLambda_{(ikl)} - \tLambda_{(jkl)} \\
	&= \dd\tLambda_{(ijkl)} + \tfrac12 \tfrac{1}{4!} \Big(  \Lambda_{(ijk)}\wedge\dd\Lambda_{(kl)} 
	+ \text{antisymmetrisation in } [ijkl] \Big)\\
	&\qquad \text{on $U_i\cap U_j \cap U_k \cap U_{l}$}, \\
		\text{etc.} \hspace{15pt} &
\end{aligned}
\end{equation}
with the final cocycle condition defined
on a octuple intersection $U_{i_1}\cap\dots\cap
U_{i_8}$. Note that this gives a generalisation of the conventional gerbe
structure, where the $\tLambda_{(ij)}$ connective structure depends on the
$\Lambda_{(ij)}$ gerbe, ultimately reflecting the Chern--Simons coupling in
eleven-dimensional supergravity~\cite{CJLP}. 

The bundle $E$ encodes all the topological information of the
supergravity background: the twisting of the tangent space $TM$ as
well as that of the gerbes, which encode the topology of the
supergravity form-field potentials.


\subsubsection{Generalised $\Edd\times\bbR^+$ structure bundle and split
  frames} 
\label{sec:gen-structure}

In all dimensions\footnote{In fact the $d\leq2$ cases essentially reduce to normal Riemannian geometry, so in what follows we will always take  $d\geq 3$.} $d\leq7$ the fibre $E_x$ of the generalised vector
bundle at $x\in M$ forms a representation space of
$\Edd\times\bbR^+$~\cite{chris,PW}. These are listed in
table~\ref{tab:gen-tang}. They correspond to the set of U-dual momentum and brane central
charges in the corresponding dimensionally reduced theories~\cite{HT},
and also appear in the dimensional reduction of West's $E_{11}$
theory~\cite{west-charges}. 

 As we discuss below, the explicit action is
defined using the $\GL(d,\bbR)$ subgroup that acts on the component
spaces $T_xM$, $\Lambda^2T^*_xM$, $\Lambda^5T^*_xM$ and
$T^*_xM\otimes\Lambda^7T^*_xM$. Note that without the additional
$\bbR^+$ action, sections of $E$ would transform as tensors weighted
by a power of $\det T^*M$. Thus it is key to extend the action to
$\Edd\times\bbR^+$ in order to define $E$ directly as the
extension~\eqref{eq:twistE}. 
\begin{table}[htb]
\begin{center}
\begin{tabular}{ll}
   $\Edd$ group & $\Edd\times\bbR^+$ rep.\\
   \hline
   $E_{7(7)}$ & $\rep{56}_\rep{1}$ \\
   $E_{6(6)}$ & $\rep{27}'_\rep{1}$ \\
   $E_{5(5)}\simeq\Spin(5,5)$  & $\rep{16}^c_\rep{1}$ \\
   $E_{4(4)}\simeq\SL(5,\bbR)$ & $\rep{10}'_\rep{1}$ \\
   $E_{3(3)}\simeq\SL(3,\bbR)\times\SL(2,\bbR)$ 
      & $(\rep{3}',\rep{2})_\rep{1}$
\end{tabular}
\end{center}
\caption{Generalised tangent space representations where the subscript
  denotes the $\bbR^+$ weight, where $\rep{1}_\rep{1}\simeq(\det{T^*M})^{1/(9-d)}$\label{tab:gen-tang}}
\end{table}

Crucially, the patching defined in~\eqref{eq:Epatch} is compatible
with this $\Edd\times\bbR^+$ action. This means that one can define a
generalised structure bundle as a sub-bundle of the frame bundle $F$
for $E$. Let $\{\hat{E}_A\}$ be a basis for $E_x$, where the label $A$ runs
over the dimension $n$ of the generalised tangent space as listed in 
table~\ref{tab:gen-tang}. The frame bundle $F$ formed from all such
bases is, by construction, a $\GL(n,\bbR)$ principal bundle. We can
then define the generalised structure bundle as the natural
$\Edd\times\bbR^+$ principal sub-bundle of $F$ compatible with the
patching~\eqref{eq:Epatch} as follows.   

Let $\hat{e}_a$ be a basis for $T_xM$ and $e^a$ the dual basis for
$T^*_xM$. We can use these to construct an explicit basis of $E_x$ as
\begin{equation}
\label{eq:e-basis}
   \{ \hat{E}_A \} = \{ \hat{e}_a \} \cup \{ e^{ab} \}  
        \cup \{ e^{a_1\dots a_5} \} \cup \{ e^{a,a_1\dots a_7} \}  , 
\end{equation}
where $e^{a_1\dots a_p}= e^{a_1}\wedge\dots\wedge e^{a_p}$ and
$e^{a,a_1\dots a_7}=e^a\otimes e^{a_1}\wedge\dots\wedge e^{a_7}$. A
generic section of $E$ at $x\in U_i$ takes the form
\begin{equation}
   V = V^A \hat{E}_A 
      = v^a \hat{e}_a + \tfrac{1}{2} \omega_{ab} e^{ab} 
         + \tfrac{1}{5!}\sigma_{a_1\dots a_5} e^{a_1\dots a_5} 
         + \tfrac{1}{7!}\tau_{a,a_1\dots a_7} e^{a,a_1\dots a_7} .
\end{equation}
As usual, a choice of coordinates on $U_i$ defines a particular such
basis where $\{\hat{E}_A\}=\{\der/\der x^m\}\cup\{\dd x^m\wedge\dd
x^n\}\cup\dots$. We will denote the components of $V$ in such a
coordinate frame by an index $M$, namely $V^M = (v^m,\omega_{mn},
\sigma_{m_1\dots m_5},\tau_{m,m_1\dots m_7})$. 

We then define a \emph{$\Edd\times\bbR^+$ basis} as one related
to~\eqref{eq:e-basis} by an $\Edd\times\bbR^+$ transformation
\begin{equation}
\label{eq:Mdef}
   V^A \mapsto V^{\prime A} = M^A{}_B V^B , \qquad
   \hat{E}_A\mapsto \hat{E}'_A=\hat{E}_B(M^{-1})^B{}_A ,
\end{equation}
where the explicit action of $M$ is defined in
appendix~\ref{app:Edd-embed}. The action has a $\GL(d,\bbR)$ subgroup 
that acts in a conventional way on the bases $\hat{e}_a$, $e^{ab}$
etc, and includes the patching
transformation~\eqref{eq:Epatch}\footnote{In analogy to the definitions
  for $O(d,d)\times\bbR^+$ generalised geometry~\cite{CSW1}, we could
  equivalently define an $\Edd\times\bbR^+$ basis using invariants
  constructed from sections of $E$. For example, in $d=7$ there is a
  natural symplectic pairing and symmetric quartic invariant that can
  be used to define~$\E7$ (in the context of generalised geometry
  see~\cite{PW}). However, these invariants differ in different
  dimension $d$ so it is more useful here to define $\Edd$ by an
  explicit action.}.

The fact that the definition of the $\Edd\times\bbR^+$ action is
compatible with the patching means that we can then define the
\emph{generalised $\Edd\times\bbR^+$ structure bundle} $\tilde{F}$
as a sub-bundle of the frame bundle for $E$ given by  
\begin{equation}
\label{eq:gen-fb}
   \tilde{F} = \big\{ (x,\{\hat{E}_A\}) : \text{$x\in M$, and 
        $\{\hat{E}_A\}$  is an $\Edd\times\bbR^+$ basis of $E_x$} 
      \big\} . 
\end{equation}
By construction, this is a principal bundle with fibre
$\Edd\times\bbR^+$. The bundle $\tilde{F}$ is the direct analogue of the frame bundle of conventional differential geometry, with $\Edd\times\bbR^+$ playing the role of $GL(d,\bbR)$.

A special class of $\Edd\times\bbR^+$ frames are those defined by a
splitting of the generalised tangent space $E$, that is, an
isomorphism of the form~\eqref{eq:Eiso}. Let $A$ and $\tA$ be three-
and six-form (gerbe) connections patched on $U_i\cap U_j$ by 
\begin{equation}
\begin{aligned}
   A_{(i)} &= A_{(j)} + \dd\Lambda_{(ij)} , \\
   \tA_{(i)} &= \tA_{(j)} + \dd\tLambda_{(ij)} 
       - \tfrac{1}{2}\dd\Lambda_{(ij)}\wedge A_{(j)} .
\end{aligned}
\end{equation}
Note that from these one can construct the globally defined field strengths
\begin{equation}
\label{eq:def-F}
\begin{aligned}
   F &= \dd A_{(i)} , \\
   \tF &= \dd \tA_{(i)} - \tfrac12 A_{(i)} \wedge F .
\end{aligned}
\end{equation}
Given a generic basis $\{\hat{e}_a\}$ for $TM$ with $\{e^a\}$ the dual
basis on $T^*M$ and a scalar function $\Delta$, we define a 
\emph{conformal split frame} $\{\hat{E}_A\}$ for $E$ by  
\begin{equation}
\label{eq:geom-basis}
\begin{aligned}
   \hat{E}_a &= \ee^{\Delta} \Big( \hat{e}_a + i_{\hat{e}_a} A
      + i_{\hat{e}_a}\tA 
      + \tfrac{1}{2}A\wedge i_{\hat{e}_a}A 
      \\ & \qquad \qquad 
      + jA\wedge i_{\hat{e}_a}\tA 
      + \tfrac{1}{6}jA\wedge A \wedge i_{\hat{e}_a}A \Big) , \\
   \hat{E}^{ab} &= \ee^\Delta \left( e^{ab} + A\wedge e^{ab} 
      - j\tA\wedge e^{ab}
      + \tfrac{1}{2}jA\wedge A \wedge e^{ab} \right) , \\
   \hat{E}^{a_1\dots a_5} &= \ee^{\Delta} \left( e^{a_1\dots a_5} 
      + jA\wedge e^{a_1\dots a_5} \right) , \\
   \hat{E}^{a,a_1\dots a_7} &= \ee^\Delta e^{a,a_1\dots a_7} , 
\end{aligned}
\end{equation}
while a \emph{split frame} has the same form but with
$\Delta=0$. To see that $A$ and $\tA$ define an
isomorphism~\eqref{eq:Eiso} note that, in the conformal split
frame, 
\begin{equation}
\label{eq:split-iso}
\begin{aligned}
   V^{(A,\tA,\Delta)}
       &= \ee^{-\Delta}\ee^{-A_{(i)}-\tA_{(i)}}V_{(i)} \\
       &= v^a \hat{e}_a + \tfrac{1}{2} \omega_{ab} e^{ab} 
         + \tfrac{1}{5!}\sigma_{a_1\dots a_5} e^{a_1\dots a_5} 
         + \tfrac{1}{7!}\tau_{a,a_1\dots a_7} e^{a,a_1\dots a_7} \\
       &\in\Gs{TM \oplus \Lambda^2T^*M \oplus \Lambda^5T^*M
         \oplus (T^*M \otimes\Lambda^7T^*M)} ,
\end{aligned}
\end{equation}
since the patching implies $\ee^{-A_{(i)}-\tA_{(i)}}V_{(i)}
=\ee^{-A_{(j)}-\tA_{(j)}}V_{(j)}$ on $U_i\cap U_j$. 

The class of split frames defines a sub-bundle of $\tilde{F}$
\begin{equation}
\label{eq:split-fb}
   P_{\text{split}} = \big\{ (x,\{\hat{E}_A\}) : \text{$x\in M$, and 
        $\{\hat{E}_A\}$ is split frame} 
      \big\} \subset \tilde{F} . 
\end{equation}
Split frames are related by transformations~\eqref{eq:Mdef} where $M$ takes
the form $M=\ee^{a+\ta}m$ with $m\in\GL(d,\bbR)$. The action of
$a+\ta$ shifts $A\mapsto A+a$ and $\tA\mapsto\tA+\ta$. This forms a
parabolic subgroup $\Ggeom=\GL(d,\bbR)\ltimes\text{$(a+\ta)$-shifts}  
\subset\Edd\times\bbR^+$ where $\text{$(a+\ta)$-shifts}$ is the
nilpotent group of order two formed of elements $M=\ee^{a+\ta}$. 
Hence $P_{\text{split}}$ is a $\Ggeom$ principal sub-bundle of
$\tilde{F}$, that is a $\Ggeom$-structure. This reflects the fact that
the patching elements in the definition of $E$ lie only in
this subgroup of $\Edd\times\bbR^+$.


\subsubsection{Generalised tensors} 
\label{sec:gen-tensor}

Generalised tensors are simply sections of vector bundles
constructed from the generalised structure bundle using different
representations of $\Edd\times\bbR^+$. We have already discussed the
generalised tangent space $E$. There are four other vector bundles
which will be of particular importance in the following. The relevant
representations are summarised in
table~\ref{tab:gen-tensors}\footnote{Note that these representations
   have already appeared in both the dimensional reduction of the
   $E_{11}$ theory~\cite{west-charges} and the tensor hierarchy
   formulation of gauged supergravity~\cite{de-form,hierarchy}.}
\begin{table}[htb]
\begin{center}
\begin{tabular}{lllll}
    dimension & $E^*$ & $\adj{\tilde{F}}\subset E\otimes E^*$ 
      & $N\subset S^2E$ & $K\subset E^*\otimes\adj{\tilde{F}}$ \\
   \hline
   7 & $\rep{56}_\rep{-1}$ 
      & $\rep{133}_\rep{0}+\rep{1}_\rep{0}$ 
      & $\rep{133}_\rep{+2}$
      & $\rep{912}_\rep{-1}$  \\
   6 & $\rep{27}_\rep{-1}$ 
      & $\rep{78}_\rep{0}+\rep{1}_\rep{0}$
      & $\rep{27}'_\rep{+2}$
      & $\rep{351}'_\rep{-1}$  \\ 
   5 & $\rep{16}^c_\rep{-1}$ 
      & $\rep{45}_\rep{0}+\rep{1}_\rep{0}$ 
      & $\rep{10}_\rep{+2}$  
      & $\rep{144}^c_\rep{-1}$  \\ 
   4 & $\rep{10}_\rep{-1}$ 
      & $\rep{24}_\rep{0}+\rep{1}_\rep{0}$ 
      & $\rep{5}'_\rep{+2}$ 
      & $\rep{40}_\rep{-1}+\rep{15}'_\rep{-1}$  \\ 
   3  & $(\rep{3},\rep{2})_\rep{-1}$ 
      & $(\rep{8},\rep{1})_\rep{0}+(\rep{1},\rep{3})_\rep{0}
         +\rep{1}_\rep{0}$ 
      & $\repp{3'}{1}_\rep{+2}$ 
      & $(\rep{3'},\rep{2})_\rep{-1}+(\rep{6},\rep{2})_\rep{-1}$ 
\end{tabular}
\end{center}
\caption{Some generalised tensor bundles\label{tab:gen-tensors}}
\end{table}

The first is the dual generalised tangent space
\begin{equation}
\label{eq:E*}
   E^* \simeq T^*M \oplus \Lambda^2TM \oplus \Lambda^5TM
         \oplus (TM \otimes\Lambda^7TM) .
\end{equation}
Given a basis $\{\hat{E}_A\}$ for $E$ we have a dual basis $\{E^A\}$ on
$E^*$ and sections of $E^*$ can be written as $Z=Z_A E^A$. 

Next we then have the adjoint bundle $\adj{\tilde{F}}$ associated with the
$\Edd\times\bbR^+$ principal bundle $\tilde{F}$
\begin{equation}
\label{eq:adj}
   \adj{\tilde{F}} \simeq \bbR \oplus \left(TM\otimes T^*M\right) 
        \oplus \Lambda^3T^*M \oplus \Lambda^6T^*M 
        \oplus \Lambda^3TM \oplus \Lambda^6TM . 
\end{equation}
By construction $\adj{\tilde{F}}\subset E\otimes E^*$ and hence we can
write sections as $R=R^A{}_B \hat{E}_A\otimes E^B$. We write the
projection on the adjoint representation as 
\begin{equation}
   \label{eq:oadj}
   \oadj : E^* \otimes E \to \adj{\tilde{F}} .
\end{equation}
It is given explicitly in~\eqref{eq:EE*-adj}. 

We also consider the sub-bundle of the symmetric product of two
generalised tangent bundles $N\subset S^2E$, 
\begin{equation}
\label{eq:N}
\begin{aligned}
   N &\simeq 
       T^*M 
       \oplus \Lambda^4T^*M
       \oplus (T^*M\otimes \Lambda^6T^*M)
       \\ & \qquad
       \oplus (\Lambda^3T^*M\otimes\Lambda^7T^*M)
       \oplus (\Lambda^6T^*M\otimes\Lambda^7T^*M). 
\end{aligned}
\end{equation}
We can write sections as $Y=Y^{AB} \hat{E}_A\otimes \hat{E}_B$ with the
projection
\begin{equation}
   \label{eq:on}
   \on : E \otimes E \to N . 
\end{equation}
It is given explicitly in~\eqref{eq:EE-N}. 

Finally, we also need the higher dimensional representation $K\subset
E^*\otimes\adj{\tilde{F}}$ listed in the last column of 
table~\ref{tab:gen-tensors}. Decomposing under 
$\GL(d,\bbR)$ one has 
\begin{equation}
\label{eq:K}
\begin{aligned}
   K &\simeq 
       T^*M 
       \oplus S^2TM 
       \oplus \Lambda^2TM
       \oplus (\Lambda^2T^*M\otimes TM)_0
       \oplus (\Lambda^3TM\otimes T^*M)_0
       \\ & \qquad
       \oplus \Lambda^4T^*M
       \oplus (\Lambda^4TM\otimes TM)_0
       \oplus \Lambda^5TM
       \oplus (\Lambda^2TM\otimes \Lambda^6TM)_0
       \\ & \qquad
       \oplus \Lambda^7T^*M 
       \oplus (TM\otimes\Lambda^7TM)
       \oplus (\Lambda^7TM\otimes\Lambda^7TM)
       \\ & \qquad
       \oplus (S^2T^*M \otimes \Lambda^7TM) 
       \oplus (\Lambda^4TM\otimes \Lambda^7TM) , 
\end{aligned}
\end{equation}
where, in fact, the $\Lambda^5TM$ term is absent when $d=5$. Note also that the zero subscripts are defined such that
\begin{equation}
\begin{aligned}
   a_{mn}{}^n &= 0 , &&& 
     &\text{if $a\in\Gs{(\Lambda^2T^*M\otimes TM)_0}$} ,\\
   a^{mnp}{}_p &= 0 , &&& 
     &\text{if $a\in\Gs{(\Lambda^3TM\otimes T^*M)_0}$} ,\\
   a^{[m_1m_2m_3m_4,m_5]}&= 0 , &&& 
     &\text{if $a\in\Gs{(\Lambda^4TM\otimes TM)_0}$} ,\\
   a^{m[n_1,n_2\dots,n_7]} &= 0 , &&& 
     &\text{if $a\in\Gs{(\Lambda^2TM\otimes \Lambda^6TM)_0}$} .
\end{aligned}
\end{equation}
Since $K\subset E^*\otimes\adj{\tilde{F}}$ we can write sections as
$T=T_{A\ph{B}C}^{\ph{A}B} E^A\otimes\hat{E}_B\otimes E^C$. 

It is interesting to note that, up to symmetries of the $E_d$ Dynkin diagram, the Dynkin labels of the representations $E$ and $N$ follow patterns as $d$ varies. For each value of $d$, the Dynkin label for $E$ can be represented on the Dynkin diagram as
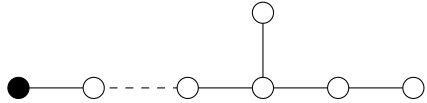
\begin{center}
\begin{tikzpicture} [dynknode/.style={circle,draw=black,fill=white, inner sep=0pt,minimum size=8}]
\draw (-1.25,0)--(-0.25,0);
\draw[dashed] (-0.25,0)--(1,0);
\draw (2,0) -- (3,0); 
\draw (1,0) -- (2,0); \draw (2,0)--(2,1);
\draw (3,0) -- (4,0);
\node at (-1.25,0) [circle,draw=black,fill=black, inner sep=0pt,minimum size=8] {}; 
\node at (-0.25,0) [dynknode] {}; 
\node at (1,0) [dynknode] {};
\node at (2,0) [dynknode] {};
\node at (3,0) [dynknode] {};
\node at (4,0) [dynknode] {};
\node at (2,1) [dynknode] {};
\end{tikzpicture}
\end{center}
while $N$ has the label
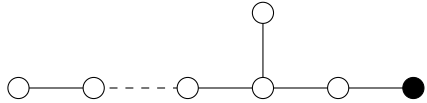
\begin{center}
\begin{tikzpicture} [dynknode/.style={circle,draw=black,fill=white, inner sep=0pt,minimum size=8}]
\draw (-1.25,0)--(-0.25,0);
\draw[dashed] (-0.25,0)--(1,0);
\draw (2,0) -- (3,0); 
\draw (1,0) -- (2,0); \draw (2,0)--(2,1);
\draw (3,0) -- (4,0);
\node at (-1.25,0) [dynknode] {}; 
\node at (-0.25,0) [dynknode] {}; 
\node at (1,0) [dynknode] {};
\node at (2,0) [dynknode] {};
\node at (3,0) [dynknode] {};
\node at (4,0) [circle,draw=black,fill=black, inner sep=0pt,minimum size=8] {};
\node at (2,1) [dynknode] {};
\end{tikzpicture}
\end{center}
%


\subsection{The Dorfman derivative and Courant bracket}
\label{sec:gen-lie}

An important property of the generalised tangent space is that it admits a generalisation of
the Lie derivative which encodes the bosonic symmetries of the
supergravity. Given $V=v+\omega+\sigma+\tau\in\Gs{E}$, one can define
an operator $\Lgen_V$ acting on any generalised tensor, which combines
the action of an infinitesimal diffeomorphism generated by $v$ and
$A$- and $\tA$-field gauge transformations generated by $\omega$ and
$\sigma$. Formally this gives $E$ the structure of a ``Leibniz
algebroid''~\cite{baraglia}. 

Acting on $V'=v'+\omega'+\sigma'+\tau'\in\Gs{E}$, one defines the
\emph{Dorfman derivative}\footnote{
  The corresponding
  object on a Courant algebroid, where the generalised structure is
  $O(d,d)$ is known as the Dorfman bracket and,
  following~\cite{baraglia}, we use the same nomenclature in this
  case too.}
or ``generalised Lie derivative''
\begin{equation}
\label{eq:Lgen}
\begin{aligned} 
   \Lgen_V V' &= \mathcal{L}_v v' 
       + \left( \mathcal{L}_v \omega' - i_{v'} \dd\omega \right)
       + \left( \mathcal{L}_v \sigma' - i_{v'} \dd\sigma
          - \omega'\wedge\dd\omega \right) 
       \\ & \qquad 
       + \left( \mathcal{L}_v \tau'
          - j\sigma'\wedge\dd\omega
          - j\omega'\wedge\dd\sigma \right) .
\end{aligned}
\end{equation}
Defining the action on a function $f$ as simply $\Lgen_V
f=\mathcal{L}_vf$, one can then extend the notion of Dorfman
derivative to a derivative on the space of $\Edd\times\bbR^+$ tensors
using the Leibniz property.

To see this, first note that we can rewrite~\eqref{eq:Lgen} in a more
$\Edd\times\bbR^+$ covariant way, in analogy with the corresponding
expressions for the conventional Lie derivative and the Dorfman
derivative in $O(d,d)\times\bbR^+$ generalised
geometry~\cite{CSW1}. One can embed the action of the partial
derivative operator via the map $T^*M\to E^*$ defined by the dual of the
exact sequences~\eqref{eq:twistE}. In coordinate indices $M$, as viewed
as mapping to a section of $E^*$, one defines   
\begin{equation}
\label{eq:d-def}
   \der_M 
      = \begin{cases} \der_m  & \text{for $M=m$} \\
         0 & \text{otherwise}
         \end{cases} . 
\end{equation}
Such an embedding has the property that under the projection onto $N^*$
we have 
\begin{equation}
   \der f \proj{N^*}\der g = 0 ,
\end{equation}
for arbitrary functions $f,g$. We will comment on this observation in section~\ref{sec:jacobi}.

One can then rewrite~\eqref{eq:Lgen} in terms of generalised
objects as
\begin{equation}
\label{eq:Lgen-cov}
   \Lgen_V V^{\prime M}
      = V^N \der_N V^{\prime M} 
         - (\der\oadj V)^M{}_N V^{\prime N} ,
\end{equation}
where $\oadj$ denotes the projection onto $\adj{\tilde{F}}$ given
in~\eqref{eq:oadj}. Concretely, from~\eqref{eq:EE*-adj} we have 
\begin{equation}
   \der\oadj V = r + a + \ta ,
\end{equation}
where $r^m{}_n=\der_n v^m$, $a=\dd\omega$ and $\ta=\dd\sigma$. We see
that the action actually lies in the adjoint of the
$\Ggeom\subset\Edd\times\bbR^+$ group. This form of the Dorfman
derivative can then be naturally extended to an arbitrary
$\Edd\times\bbR^+$ tensor by taking that appropriate adjoint action on
the $\Edd\times\bbR^+$ representation.  

Note that we can also define a bracket by taking the
antisymmetrisation of the Dorfman derivative. This was originally
given in~\cite{PW} where it was called the ``exceptional Courant
bracket'', and re-derived in~\cite{baraglia}. It is given by  
\begin{equation}
\label{eq:ECB}
\begin{split}
   \Bgen{V}{V'} &= \tfrac{1}{2}\left( \Lgen_V V' - \Lgen_{V'} V \right) \\
     &= [v,v']
        + \mathcal{L}_v\omega' - \mathcal{L}_{v'}\omega
           - \tfrac{1}{2}\dd\left(i_v\omega'-i_{v'}\omega\right)
     \\ & \quad
        + \mathcal{L}_v\sigma' - \mathcal{L}_{v'}\sigma
           - \tfrac{1}{2}\dd\left(i_v\sigma'-i_{v'}\sigma\right)
        + \tfrac{1}{2}\omega\wedge\dd\omega'
           - \tfrac{1}{2}\omega'\wedge\dd\omega
     \\ & \quad
        + \tfrac{1}{2}\mathcal{L}_v\tau'
           - \tfrac{1}{2}\mathcal{L}_{v'}\tau
        + \tfrac{1}{2}\big(
           j\omega\wedge\dd\sigma' - j\sigma'\wedge\dd\omega \big)
        - \tfrac{1}{2}\big(
           j\omega'\wedge\dd\sigma - j\sigma\wedge\dd\omega' \big) .
\end{split}
\end{equation}
Note that the group generated by closed $A$ and $\tA$
shifts is a semi-direct product
$\Omega^3_{\text{cl}}(M)\ltimes\Omega^6_{\text{cl}}(M)$ and
corresponds to the symmetry group of gauge transformations in the
supergravity. The full automorphism group of the exceptional Courant
bracket is then the local symmetry group of the supergravity
$G_{\text{sugra}}=
\Diff(M)\ltimes(\Omega^3_{\text{cl}}(M)\ltimes\Omega^6_{\text{cl}}(M))$.

For $U,V,W\in\Gs{E}$, the Dorfman derivative also satisfies the Leibniz identity
\begin{equation}
\label{eq:Leibniz}
   \Lgen_U (\Lgen_V W) -  \Lgen_V (\Lgen_U W) = \Lgen_{\Lgen_U V} W ,
\end{equation}
and hence $E$ is a ``Leibniz algebroid". On first inspection, one might expect that the bracket of $\Bgen{U}{V}$ should appear on the RHS. However, the statement is correct since one can show that
\begin{equation}
\label{eq:LeibnizBracket}
    \Lgen_{\Bgen{U}{V}} W = \Lgen_{\Lgen_U V} W ,
\end{equation}
so that the RHS is automatically antisymmetric in $U$ and $V$.


\subsection{Generalised $\Edd\times\bbR^+$ connections and torsion} 
\label{sec:conns}

We now turn to the definitions of generalised connections and
torsion. Generalised connections on algebroids were first introduced
by Alekseev and Xu~\cite{AX,CSX}. To study the dynamics of $E_{7(7)}$
geometries with an eleven-dimensional supergravity origin and
supersymmetric backgrounds, related notions were also developed
by~\cite{hillmann,GO}. Here, for the $\Edd\times\bbR^+$ case, we
follow much the same procedure and conventions as in~\cite{CSW1},
where we gave the precise definitions relevant for type II
supergravity, taking care to include an $\bbR^+$ factor in the
generalised structure bundle. 


\subsubsection{Generalised connections}
\label{sec:conn-def}

We first define generalised connections that are compatible with the
$\Edd\times\bbR^+$ structure. These are first-order linear differential
operators $\Dgen$, such that, given $W\in\Gs{E}$, in frame indices, 
\begin{equation}
   \Dgen_M W^A = \der_M W^A + \Omega_M{}^A{}_B W^B . 
\end{equation}
where $\Omega$ is a section of $E^*$ (denoted by the $M$ index) taking
values in $\Edd\times\bbR^+$ (denoted by the $A$ and $B$ frame
indices), and as such, the action of $\Dgen$ then extends naturally to any
generalised $\Edd\times\bbR^+$ tensor. 

A simple example of a generalised connection can be constructed as
follows. One starts with a conventional connection $\nabla$ and a conformal
split frame of the form~\eqref{eq:geom-basis}. Given the
isomorphism~\eqref{eq:split-iso}, by construction
$v^a\hat{e}_a\in\Gs{TM}$, 
$\frac{1}{2}\omega_{ab}e^{ab}\in\Gs{\Lambda^2T^*M}$ etc and
hence $\nabla_m v^a$ and $\nabla_m \omega_{ab}$ are
well-defined. The generalised connection defined by $\nabla$ lifted to
an action on $E$ by the conformal split frame then defines a
generalised connection $D^\nabla$ as 
\begin{equation}
\label{eq:Dgen-embed}
   \Dgen^\nabla_M V
    = \begin{cases}
         \begin{aligned}
            &(\nabla_m v^a) \hat{E}_a 
            + \tfrac{1}{2} (\nabla_m \omega_{ab}) \hat{E}^{ab} 
            \\ & \quad
            + \tfrac{1}{5!} (\nabla_m \sigma_{a_1\dots a_5}) 
               \hat{E}^{a_1\dots a_5} 
            + \tfrac{1}{7!} (\nabla_m \tau_{a,a_1\dots a_7})
               \hat{E}^{a,a_1\dots a_7} 
         \end{aligned}
      & \text{for $M=m$,} \\
      0  & \text{otherwise.} 
      \end{cases}  
\end{equation}
%


\subsubsection{Generalised torsion}
\label{sec:torsion}

We define the \emph{generalised torsion} $T$ of a generalised connection
$\Dgen$ in direct analogy to the conventional definition and to the one
we defined in the $O(d,d)\times\bbR^+$ description of 
type II theories~\cite{CSW1}. 

Let $\alpha$ be any generalised $\Edd\times\bbR^+$ tensor and let
$\Lgen^\Dgen_V\alpha$ be the Dorfman derivative~\eqref{eq:Lgen-cov}
with $\der$ replaced by $\Dgen$. The generalised torsion is a linear
map  $T:E\to \adj(\tilde{F})$ defined by 
\begin{equation}
\label{eq:Tgen-def}
   T(V)\cdot \alpha 
       = \Lgen^\Dgen_V \alpha - \Lgen_V \alpha , 
\end{equation}
for any $V\in\Gs{E}$ and where $T(V)$ acts via the adjoint
representation on $\alpha$. Let $\{\hat{E}_A\}$ be an
$\Edd\times\bbR^+$ frame for $E$ and $\{E^A\}$ be the dual frame for
$E^*$ satisfying $E^A(\hat{E}_B)=\delta^A{}_B$. We then have the
explicit expression 
\begin{equation}
\label{eq:Tgen-basis}
   T(V) = V^C \left[ \Omega_{C\ph{A}B}^{\ph{C}A} 
           - \Omega_{B\ph{A}C}^{\ph{B}A}
           - E^A(\Lgen_{\hat{E}_C} \hat{E}_B)
           \right] \hat{E}_A\oadj E^B . 
\end{equation}
Note that we are projecting onto the adjoint representation on the
$A$ and $B$ indices. Note also that in a coordinate frame the last
term vanishes. 

Viewed as a generalised $\Edd\times\bbR^+$ tensor we have
$T\in\Gs{E^*\otimes\adj{\tilde{F}}}$. However, the form of the Dorfman
derivative means that fewer components actually survive and we find 
\begin{equation}
   T \in \Gs{K \oplus E^*} , 
\end{equation}
where $K$ was defined in table~\ref{tab:gen-tensors}. Note that these
representations are exactly the same ones that appear in the embedding
tensor formulation of gauged supergravities~\cite{embedT1}, including
gaugings~\cite{embedT2} of the so-called ``trombone''
symmetry~\cite{trombone}. We will comment on this further in
section~\ref{sec:embedT}. 

As an example, we can calculate the torsion of the generalised connection
$\Dgen^\nabla$ defined by a conventional connection $\nabla$ and a
conformal split frame as given in~\eqref{eq:Dgen-embed}. We find 
\begin{equation}
\label{eq:splitT}
   T(V) = \ee^\Delta \left(- i_v \dd\Delta + i_v T + v\otimes \dd\Delta
              - i_v F + \dd \Delta \wedge \omega 
              - i_v \tF + \omega \wedge F 
              + \dd \Delta \wedge \sigma \right) , 
\end{equation}
where we are using the isomorphism~\eqref{eq:adj}, $F$ and $\tF$
are the field strengths~\eqref{eq:def-F}, and
$(i_vT)^\mu{}_\nu=v^\lambda T^{\mu}{}_{\lambda\nu}$ is an element of
$TM\otimes T^*M$ with $T$ the conventional torsion of $\nabla$.  


\subsection{The ``section condition'', Jacobi identity and the absence
   of generalised curvature} 
\label{sec:jacobi}

Restricting our analysis to $d \leq 6$, we find that the bundle $N$ given in~\eqref{eq:N} measures the failure of the generalised tangent bundle to satisfy the properties of a Lie algebroid. This follows from the observation that the difference between the Dorfman derivative and the exceptional Courant bracket (that is, the symmetric part of the Dorfman derivative), for $V,V'\in \Gs{E}$, is precisely given by\footnote{For $d\geq7$ the RHS can no longer be written covariantly as a derivative of an $E_{d(d)}\times\bbR^+$ tensor built from $U$ and $V$. Similar complications occur in the discussion of the curvature below. This is the reason for the restriction to $d\leq6$ in this section.}
\begin{equation}
\label{eq:bracket-diff}
\Lgen_V V' - \Bgen{V}{V'} = \tfrac{1}{2}\dd \left( i_v \omega ' + i_{v'} \omega - i_v \sigma ' - i_{v'} \sigma + \omega \wedge \omega '\right) = \partial \proj{E} ( V \on V'),
\end{equation}
where the last equality stresses the $E_{d(d)}\times\bbR^+$ covariant form of the exact term. Therefore, while the Dorfman derivative satisfies a sort of Jacobi identity via the Leibniz identity~\eqref{eq:Leibniz}, the Jacobiator of the exceptional Courant bracket, like that of the $O(d,d)$ Courant bracket, does not vanish in general. In fact, it can be shown that
\begin{equation}
\text{Jac}(U,V,W) = \Bgen{\Bgen{U}{V}}{W} + \text{c.p.} = \tfrac{1}{3}\partial \proj{E} \left( \Bgen{U}{V} \on W + \text{c.p.} \right) ,
\end{equation}
where $U,V,W \in \Gs{E}$ and c.p. denotes cyclic permutations in $U,V$ and $W$. We see that both the failure of the exceptional Courant bracket to be Jacobi and the Dorfman derivative to be antisymmetric is measured by an exact term given by the $\on$ projection. The proof is essentially the same as the one for the $O(d,d)$ case, see for example~\cite{Gualtieri}, section \textbf{3.2}\footnote{Note that in the $O(d,d)$ case the fibre of $N$ is the $\rep{1_{+2}}$ representation, so $N$ is a trivial bundle.}.

Similarly, and as was the case with $O(d,d)\times\bbR^+$ generalised connections, for notions of generalised curvature one finds the naive definition $\BLie{\Dgen_U}{\Dgen_V}W-\Dgen_{\Bgen{U}{V}}W$ is not a tensor and its failure to be covariant is measured by the projection of the first two arguments to $N$. Explicitly, taking $U \rightarrow fU$, $V \rightarrow gV$ and  $W \rightarrow hW$ for some scalar functions $f,g,h$, we obtain
\begin{equation}
\label{eq:Dgen-comm-linear}
\begin{aligned}
   &\BLie{\Dgen_{fU}}{\Dgen_{gV}}hW - \Dgen_{\Bgen{fU}{gV}} hW 
       \\ & \qquad 
       = fgh\left( \BLie{\Dgen_U}{\Dgen_V}W - \Dgen_{\Bgen{U}{V}} W \right)
                 - \tfrac{1}{2} h \Dgen_{\left(f\partial g - g \partial f\right) \proj{E} ( U \proj{N} V ) } W . 
\end{aligned}
\end{equation}
Note, however, that it is still possible to define analogues of the
Ricci tensor and scalar when there is additional structure on the
generalised tangent space, as we see in the following section.  
 
Finally, we note that from the point of view of ``double field
theory''-like
geometries~\cite{siegel,dft,dft2,BP4d,BP5d,thompson,BP-alg,BPW}, 
the equation 
\begin{equation}
\label{eq:section-cond}
   \partial f \proj{N^*} \partial g = 0,
\end{equation}
for any functions $f$ and $g$ acquires a special interpretation. In
these theories, one starts by enlarging the spacetime manifold so that
its dimension matches that of the generalised tangent space. The
partial derivative $\der_Mf$ is then generically non-zero for all
$M$. However, the corresponding Dorfman derivative does not then
satisfy the Leibniz property, nor is the action for the generalised
metric invariant. One must instead impose a ``section condition'' or
``strong constraint''. In the original $O(d,d)$ double field
theory the condition takes the form $(\partial^Mf)( \partial_Mg) =
0$. It implies that, in fact, the fields only depend on half the
coordinates. For exceptional geometries, the $d=4$ case was thoroughly
analysed in~\cite{BP-alg}, and is given
by~\eqref{eq:section-cond}. Again it implies that the fields depend on
only $d$ of the coordinates. 

It is in fact easy to show that satisfying~\eqref{eq:section-cond}
always implies the Leibniz property. Thus it gives the section
condition in general dimension. In generalised geometry it is
satisfied identically by taking $\der_M$ of the
form~\eqref{eq:d-def}. However given the $\Edd\times\bbR^+$ covariant
form of the Dorfman derivative~\eqref{eq:Lgen-cov}, any subspace of
$E^*$ in the same orbit under $\Edd\times\bbR^+$ will also satisfy the
Leibniz condition. Note further that any such subspace, like $T^*$, is
invariant under an action of the parabolic subgroup $\Ggeom$.


\section{$\Hd$ structures and torsion-free connections}
\label{sec:Hd}

We now turn to the construction of the analogue of the Levi--Civita
connection by considering additional structure on the generalised
tangent space. Again, this closely follows the constructions in
$O(d,d)\times\bbR^+$ generalised geometry~\cite{CSW1}.  

We consider $H_d$ structures on $E$ where $H_d$ is the maximally
compact\footnote{Note that one could equally consider the non-compact versions of $\Hd$ by switching the signature of the metric in appendix~\ref{app:Hd} so that it defines an $SO(p,q)$ subgroup of $GL(d,\bbR)$, and the corresponding results then follow identically. For instance, if in $d=7$ one chooses the $SO(6,1)$ signature, one would obtain the non-compact  $SU^*(8)$ subgroup of $E_{7(7)}\times\bbR^+$, which would be relevant for discussing timelike reductions of 11-dimensional supergravity~\cite{Hull:1998br}.} subgroup of $\Edd$. These, or rather their double covers\footnote{We give the double covers of the maximally compact
group, since we will be interested in the analogues of spinor
representations. A necessary and sufficient condition for the existence of the double cover is the vanishing of the 2nd Stiefel-Whitney class of the generalised tangent bundle~\cite{chris}. As the underlying manifold is spin by assumption, this is automatically satisfied.} $\dHd$ are listed in table~\ref{tab:coset}. We will then be interested in
generalised connections $\Dgen$ that preserve the $H_d$ structure. We find it is always possible
to construct torsion-free connections of this type but they are not
unique. Nonetheless we show that, using the $H_d$ structure, one can
construct unique projections of $\Dgen$, and that these can be used to
define analogues of the Ricci tensor and scalar curvatures with a
local $H_d$ symmetry. 
\begin{table}[htb]
\begin{center}
\begin{tabular}{lll}
   $\Edd$ group & $\dHd$ group & $\adj{P}^\perp = \adj{\tilde{F}}/\adj{P}$ \\
   \hline
   $E_{7(7)}$ & $\SU(8)$ & $\rep{35}+\rep{\bar{35}}+\rep{1}$ \\
   $E_{6(6)}$ & $\USp(8)$ & $\rep{42}+\rep{1}$ \\
   $E_{5(5)}\simeq\Spin(5,5)$  & $\Spin(5)\times\Spin(5)$ 
      & $\repp{5}{5}+\repp{1}{1}$ \\
   $E_{4(4)}\simeq\SL(5,\bbR)$ & $\Spin(5)$ & $\rep{14}+\rep{1}$ \\
   $E_{3(3)}\simeq\SL(3,\bbR))\times\SL(2,\bbR)$ 
      & $\Spin(3)\times\Spin(2)$ 
      & $\repp{5}{1}+\repp{1}{2} + \repp{1}{1}$
\end{tabular}
\end{center}
\caption{Double covers of the maximal compact subgroups of $\Edd$ and
  $\Hd$ representations of the coset bundle\label{tab:coset}}
\end{table}
%


\subsection{$\Hd$ structures and the generalised metric}
\label{sec:OdOd}

An $H_d$ structure on the tangent space is a set of frames related by
$H_d$ transformations. This is the direct analogy of metric structure,
where one considers the set of orthonormal frames related by $O(d)$
transformations. Formally it defines an $H_d$ principal sub-bundle of
the generalised structure bundle $\tilde{F}$, that is 
\begin{equation}
   \label{eq:Hd-structure}
   P \subset \tilde{F} \text{ with fibre $H_d$}.
\end{equation}
The choice of such a structure is parametrised, at each point on the
manifold, by an element of the coset $(\Edd\times\bbR^+)/\Hd$. The
corresponding representations are listed in
table~\ref{tab:coset}. Note that there is always a singlet
corresponding to the $\bbR^+$ factor. 

One can construct elements of $P$ concretely, that is, identify the
analogues of ``orthonormal'' frames, in the following way. Given an
$H_d$ structure, it is always possible to put the $H_d$ frame in a
conformal split form, namely,   
\begin{equation}
\label{eq:Hdsplit}
\begin{aligned}
   \hat{E}_a &= \ee^{\Delta} \Big( \hat{e}_a + i_{\hat{e}_a} A
      + i_{\hat{e}_a}\tA 
      + \tfrac{1}{2}A\wedge i_{\hat{e}_a}A 
      \\ & \qquad \qquad 
      + jA\wedge i_{\hat{e}_a}\tA 
      + \tfrac{1}{6}jA\wedge A \wedge i_{\hat{e}_a}A \Big) , \\
   \hat{E}^{ab} &= \ee^\Delta \left( e^{ab} + A\wedge e^{ab} 
      - j\tA\wedge e^{ab}
      + \tfrac{1}{2}jA\wedge A \wedge e^{ab} \right) , \\
   \hat{E}^{a_1\dots a_5} &= \ee^{\Delta} \left( e^{a_1\dots a_5} 
      + jA\wedge e^{a_1\dots a_5} \right) , \\
   \hat{E}^{a,a_1\dots a_7} &= \ee^\Delta e^{a,a_1\dots a_7} . 
\end{aligned}
\end{equation}
Any other frame is then related by an $H_d$ transformation of the form
given in appendix~\ref{app:Hd}. Concretely given
$V=V^A\hat{E}_A\in\Gs{E}$ expanded in such a frame, different frames
are related by 
\begin{equation}
   V^A \mapsto V^{\prime A} = H^A{}_B V^B , \qquad
   \hat{E}_A\mapsto \hat{E}'_A=\hat{E}_B(H^{-1})^B{}_A ,
\end{equation}
where $H$ is defined in~\eqref{eq:Haction}. Note that the
$O(d)\subset\Hd$ action simply rotates the $\hat{e}_a$
basis, defining a set of orthonormal frames for a conventional
metric $g$. It also keeps the frame in the conformal split form. Thus
the set of conformal split $\Hd$ frames actually forms an $O(d)$
structure on $E$, that is 
\begin{equation}
   \label{eq:Hd-structure-split}
   (P \cap P_{\text{split}}) \subset \tilde{F} 
      \text{ with fibre $O(d)$} .
\end{equation}

One can also define the generalised metric acting on $V\in\Gs{E}$ as 
\begin{equation}
    G(V,V) = v^2 + \tfrac{1}{2!}\omega^2 
      + \tfrac{1}{5!}\sigma^2 + \tfrac{1}{7!}\tau^2,
\end{equation}
$v^2=v_av^a$, $\omega^2=\omega_{ab}\omega^{ab}$, etc as
in~\eqref{eq:*norm}, are
evaluated in an $\Hd$ frame and indices are contracted using the
flat frame metric $\delta_{ab}$ (as used to define the $\Hd$ subgroup
in appendix~\ref{app:Hd}). Since, by definition, this is independent
of the choice of $\Hd$ frame, it can be evaluated in the conformal split
representative~\eqref{eq:Hdsplit}. Hence one sees explicitly that the
metric is defined by the fields $g$, $A$, $\tA$ and $\Delta$ that
determine the coset element. 

Note that the $\Hd$ structure embeds as
$\Hd\subset\Edd\subset\Edd\times\bbR^+$. This mirrors the chain of embeddings in Riemannian geometry $SO(d)\subset SL(d,\bbR)\subset GL(d,\bbR)$ which allows one to define a
$\det T^*M$ density that is $SO(d)$ invariant, $\sqrt{g}$. Likewise, here we can define a density that is $\Hd$ (and $\Edd$) invariant, corresponding to the choice of $\bbR^+$ factor which, in terms of the 
conformal split frame, is given by
\begin{equation}
\label{eq:Phi-def}
   \vol_G = \sqrt{g}\, \ee^{(9-d)\Delta} , 
\end{equation}
as can be seen from appendix~\ref{app:Edd-def}. This can also be defined as the determinant of $G$ to a suitable power.


\subsection{Torsion-free, compatible connections}
\label{sec:genLC}

A generalised connection $\Dgen$ is compatible with the $\Hd$
structure $P\subset\tilde{F}$ if
\begin{equation}
   \Dgen G = 0 , 
\end{equation}
or, equivalently, if the derivative acts only in the $\Hd$ sub-bundle. In
this subsection we will show, in analogy to the construction of the
Levi--Civita connection, that 
\begin{quote}
   \textit{Given an $\Hd$ structure $P\subset\tilde{F}$
     there always exists a torsion-free, compatible generalised
     connection $\Dgen$. However, it is not unique.} 
\end{quote}
We construct the compatible connection explicitly by working in the
conformal split $\Hd$ frame~\eqref{eq:Hdsplit}. However the
connection is $\Hd$ covariant, so the form in any another frame
simply follows from an $\Hd$ transformation.  

Let $\nabla$ be the Levi--Civita connection for the metric $g$. We can
lift the connection to an action on $V\in\Gs{E}$ by defining, as
in~\eqref{eq:Dgen-embed},  
\begin{equation}
   \Dgen^\nabla_M V
    = \begin{cases}
         \begin{aligned}
            &(\nabla_m v^a) \hat{E}_a 
            + \tfrac{1}{2} (\nabla_m \omega_{ab}) \hat{E}^{ab} 
            \\ & \quad
            + \tfrac{1}{5!} (\nabla_m \sigma_{a_1\dots a_5}) 
               \hat{E}^{a_1\dots a_5} 
            + \tfrac{1}{7!} (\nabla_m \tau_{a,a_1\dots a_7})
               \hat{E}^{a,a_1\dots a_7} 
         \end{aligned}
      & \text{for $M=m$,} \\
      0  & \text{otherwise.} 
      \end{cases}  
\end{equation}
Since $\nabla$ is compatible with the $O(d)\subset\Hd$ subgroup, it is
necessarily an $\Hd$-compatible connection. However, $\Dgen^\nabla$ is
not torsion-free. From~\eqref{eq:splitT}, since $\nabla$ is
torsion-free (in the conventional sense), we have  
\begin{equation}
   T(V) = \ee^\Delta \left( -i_v \dd\Delta + v\otimes \dd\Delta
              - i_v F + \dd \Delta \wedge \omega 
              - i_v \tilde{F} + \omega \wedge F 
              + \dd \Delta \wedge \sigma \right). 
\end{equation}

To construct a torsion-free compatible connection we simply modify
$\Dgen^\nabla$. A generic generalised connection $\Dgen$ can always
be written as  
\begin{equation}
   \Dgen_M W^ A = \Dgen^\nabla_M W^A + \Sigma_M{}^A{}_B W^B . 
\end{equation}
If $\Dgen$ is compatible with the $\Hd$ structure then
\begin{equation}
   \Sigma \in \Gs{E^*\otimes\adj{P}} ,
\end{equation}
that is, it is a generalised covector taking values in the adjoint of
$\Hd$. The problem is then to find a suitable $\Sigma$ such that the
torsion of $\Dgen$ vanishes. Fortunately, decomposing under $\Hd$ one
finds that all the representations that appear in the torsion are
already contained in $\Sigma$. Thus a solution always exists, but is
not unique\footnote{In $d=3$ all the components of $\Sigma$ are contained in the torsion representations, $E^*\otimes\adj{P} \simeq K\oplus E^*$, and so, in that particular case, the generalised connection is in fact completely determined.}. The relevant representations are listed in
table~\ref{tab:U}. As $\Hd$ tensor bundles one has  
\begin{equation}
   E^*\otimes\adj{P} \simeq (K\oplus E^*) \oplus U ,
\end{equation}
so that the torsion $T\in\Gs{K\oplus E^*}$ and the unconstrained part
of $\Sigma$ is a section of $U$. 
\begin{table}[htb]
\begin{center}
\begin{tabular}{lll}
   dimension & $K\oplus E^*$ 
   & $U \simeq (E^*\otimes \adj{P}) / (K\oplus E^*)$  \\  
   \hline
   7 & $\rep{28} + \bar{\rep{28}} + \rep{36} + \bar{\rep{36}}+
           \rep{420} + \bar{\rep{420}}$ 
        & $\rep{1280}+ \bar{\rep{1280}}$  \\
   6 & $\rep{27} + \rep{36} + \rep{315}$
        & $\rep{594}$ \\
   5 & $\repp{4}{4} + \repp{4}{4}+ \repp{16}{4} + \repp{4}{16}$
        & $\repp{20}{4} + \repp{4}{20} $ \\
   4 & $\rep{1} + \rep{5} + \rep{10} + \rep{14} + \rep{35'}$
        & $\rep{35}$ \\
   3 & $\repp{1}{2} + \repp{3}{2} + \repp{3}{2} + \repp{5}{2}$
        & - 
\end{tabular}
\end{center}
\caption{Components of the connection $\Sigma$ that are constrained by
  the torsion, $T$, and the unconstrained ones, $U$, as $\Hd$
  representations\label{tab:U}} 
\end{table}

The solution for $\Sigma$ can be written very explicitly as
follows. Contracting with $V\in\Gs{E}$ so $\Sigma(V)\in\adj{P}$ and
using the basis for the adjoint of $\Hd$ given
in~\eqref{eq:Hd-alg} and~\eqref{eq:Hd-embed} we have 
\begin{equation}
\begin{aligned}
   \Sigma(V)_{ab} &= \ee^\Delta \left( 2 \left( \tfrac{7-d}{d-1} \right) v_{[a} \der_{b]} \Delta
		+ \tfrac{1}{4!} \omega_{cd} F^{cd}{}_{ab} 
		+ \tfrac{1}{7!} \sigma_{c_1 \dots c_5} 
                    \tilde{F}^{c_1 \dots c_5}{}_{ab} 
                + C(V)_{ab} \right) ,\\
   \Sigma(V)_{abc} &= \ee^\Delta \left( \tfrac{6}{(d-1)(d-2)} (\dd \Delta \wedge \omega)_{abc} 
   	+ \tfrac14 v^d F_{dabc} + C(V)_{abc} \right) ,\\
   \Sigma(V)_{a_1\dots a_6} 
      &= \ee^\Delta \left( \tfrac{1}{7} v^b \tilde{F}_{ba_1\dots a_6} + C(V)_{a_1\dots a_6} \right) , 
\end{aligned}
\end{equation}
where the ambiguous part of the connection
$\am\in\Gs{E^*\otimes\adj{P}}$ projects to zero under the map to the
torsion representation $K\oplus E^*$, that is 
\begin{equation}
   \label{eq:am-cond}
   \am \in \Gs{U} .
\end{equation}
Using the embedding of $\dHd$ in
$\Cliff(d;\bbR)$ given in~\eqref{eq:Hd-cliff} we can thus write the
full connection as
\begin{equation}
\label{eq:Dgen-sol}
\begin{aligned}
   \Dgen_a 
      &= \ee^\Delta \left( \nabla_a
          + \tfrac{1}{2} {\left( \tfrac{7-d}{d-1} \right)} (\der_b \Delta) \gamma_a{}^b
          - \tfrac12 \tfrac{1}{4!} F_{ab_1 b_2 b_3} \gamma^{b_1 b_2 b_3}
          - \tfrac12 \tfrac{1}{7!} \tF_{ab_1\dots b_6}\gamma^{b_1\dots b_6}
          + \slashed{\am}_a \right), \\
   \Dgen^{a_1 a_2} 
      &= \ee^\Delta \left( \tfrac{1}{4} \tfrac{2!}{4!} F^{a_1 a_2}{}_{b_1 b_2} \gamma^{b_1 b_2}
      	-  \tfrac{3}{(d-1)(d-2)} (\der_b \Delta) \gamma^{a_1 a_2 b}
          + \slashed{\am}^{a_1 a_2} \right), \\
   \Dgen^{a_1\dots a_5} 
      &= \ee^\Delta \left( \tfrac{1}{4} \tfrac{5!}{7!} \tF^{a_1\dots a_5}{}_{b_1b_2}\gamma^{b_1b_2}
          + \slashed{\am}^{a_1\dots a_5}\right) , \\
   \Dgen^{a,a_1\dots a_7} 
      &= \ee^\Delta \left( \slashed{\am}^{a,a_1\dots a_7} \right) ,
\end{aligned}
\end{equation}
where 
\begin{equation}
\begin{aligned}
  \slashed{\am}_m 
       &= \tfrac{1}{2} \left( \tfrac{1}{2!} \am_{m,ab} \gamma^{ab}
          - \tfrac{1}{3!} \am_{m,a_1a_2a_3} \gamma^{a_1a_2a_3}
          - \tfrac{1}{6!} \am_{m,a_1 \dots a_6}\gamma^{a_1 \dots a_6} \right)
          , \\
  \slashed{\am}^{m_1m_2} 
       &= \tfrac{1}{2} \left( \tfrac{1}{2!} \am^{m_1m_2}{}_{ab} \gamma^{ab}
          - \tfrac{1}{3!} \am^{m_1m_2}{}_{a_1a_2a_3} 
              \gamma^{a_1a_2a_3}
          - \tfrac{1}{6!} \am^{m_1m_2}{}_{a_1 \dots a_6}
              \gamma^{a_1 \dots a_6} \right)
              , \\
       & \qquad \text{etc.}
\end{aligned}
\end{equation}
is the embedding of the ambiguous part of the 
connection\footnote{It is interesting to compare this connection to
   the one defined in~\cite{GO}. There $\Sigma$ is chosen to lie
   solely in the $\rep{912}$ representation. This leads to a unique
   torsion-free connection, which is, however, not compatible with the
   generalised metric.}. 


\subsection{Unique operators and generalised $\Hd$ curvatures}
\label{sec:curv}

We now turn to the construction of unique operators and curvatures from the torsion-free and $\dHd$-compatible connection $\Dgen$ constructed in the previous section. To keep the $\dHd$ covariance manifest in all dimensions, we will necessarily have to maintain the discussion in this section fairly abstract. We should note, however, that the entire construction can be made very concrete. In~\cite{CSW2} we will present the details for particular dimensions, such as the $d=7$ case where the unique operators and the curvatures can be explicitly written out in $\tilde{H}_7 = SU(8)$ indices.

Given a bundle $X$ transforming as some representation of $\dHd$, we define the map 
\begin{equation}
   \mathcal{Q}_{X}: U\otimes X \longrightarrow E^* \otimes X, 
\end{equation}
via the embedding $U\subset E^*\otimes\adj{P}$ and the adjoint action
of $\adj{P}$ on $X$. We then have the projection
\begin{equation}
   \mathcal{P}_{X}: E^* \otimes X \longrightarrow \frac{E^* \otimes
     X}{\text{Im}\mathcal{Q}_{X}}. 
\end{equation}
Recall that the ambiguous part $\am$ of the connection $\Dgen$ is
a section of $U$, which acts on $X$ via the map $\mathcal{Q}_X$. If
$\alpha\in\Gs{X}$, then, by construction,
$\mathcal{P}_{X}(\Dgen\otimes\alpha)$ is uniquely defined, independent
of $\am$. 

We can construct explicit examples of such operators as
follows. Consider two real $\dHd$ bundles $S$ and $J$, which we 
refer to as the ``spinor'' bundle and the ``gravitino''
bundle respectively, since the supersymmetry parameter and the
gravitino field in supergravity are sections of them. The
relevant $\dHd$ representations are listed in table~\ref{tab:SJ}. 
\begin{table}[htb]
\begin{center}
\begin{tabular}{lll}
	$\dHd$ & $S$ & $J$ \\ 
	\hline
	$\SU(8)$ & $\rep{8}+\bar{\rep{8}}$ & $\rep{56}+\bar{\rep{56}}$ \\
	$\USp(8)$ & $\rep{8}$ & $\rep{48}$\\
	$\USp(4)\times U\! Sp(4)$ & $\repp{4}{1} + \repp{1}{4}$ 
            & $\repp{4}{5} + \repp{5}{4}$\\
	$\USp(4)$ & $\rep{4}$ & $\rep{16}$	\\
	$\SU(2) \times U(1)$ & $\rep{2}_{\rep{1}} + \rep{2}_{\rep{-1}}$ 
            & $\rep{4}_{\rep{1}} + \rep{4}_{\rep{-1}} 
                + \rep{2}_{\rep{3}} + \rep{2}_{\rep{-3}}$ 
\end{tabular}
\end{center}
\caption{Spinor and gravitino representations in each dimension\label{tab:SJ}}
\end{table}
Note that the spinor representation is simply the $\Cliff(d;\bbR)$
spinor representation using the embedding~\eqref{eq:Hd-cliff}. 

One finds that under the projection $\mathcal{P}_X$ we have\footnote{Note that there is an exception for $d=3$ since, as was previously mentioned, in that case the entire metric compatible, torsion-free connection is uniquely determined, and so $\mathcal{P}_{X}$ is just the identity map and $\mathcal{P}_{X}(E^* \otimes X) = E^* \otimes X$ for any bundle $X$.}
\begin{equation}
\begin{aligned}
   \mathcal{P}_{S}(E^* \otimes S) \simeq S \oplus J , \\
   \mathcal{P}_{J}(E^* \otimes J) \simeq S \oplus J . 
\end{aligned}
\end{equation}
Therefore, for any $\varepsilon \in \Gs{S}$ and $\psi \in \Gs{J}$, one
has that the following are unique for any torsion-free connection 
\begin{equation}
\label{eq:unique-ops}
\begin{aligned}
   \Dgen \proj{J} \varepsilon, \qquad 
   \Dgen \proj{S} \varepsilon, \\ 
   \Dgen \proj{J} \psi, \qquad 
   \Dgen \proj{S} \psi, 
\end{aligned}
\end{equation}
where $\proj{X}$ denotes the projection 
onto the $X$ bundle. 

One can show that the first two expressions encode 
the supersymmetry variation of the internal and external gravitino
respectively, while the latter two are related to the gravitino
equation of motion. This will be described in more detail
in~\cite{CSW2}.

We would now like to define measures of generalised curvature. As was mentioned in section~\ref{sec:jacobi}, the natural definition of a Riemann curvature does not result in a tensor. Nonetheless, for a torsion-free, $\dHd$-compatible connection $\Dgen$ there does exist a generalised Ricci tensor $\GenRicci_{AB}$, and it is a section of the bundle 
\begin{equation}
	\adj{P}^\perp = \adj{\tilde{F}}/\adj{P} \subset E^* \otimes E^* ,
\end{equation}
where the last relation follows because, as representations of $\Hd$, $E \simeq E^*$.
It is not immediately apparent that we can make such a definition, but $\GenRicci_{AB}$ can in fact  be constructed from compositions of the unique operators~\eqref{eq:unique-ops} as 
\begin{equation}
\label{eq:GenRS}
\begin{aligned}
   \Dgen \proj{J} (\Dgen \proj{J}
         \varepsilon) 
      + \Dgen \proj{J} (\Dgen
         \proj{S} \varepsilon) 
      &= \GenRic \cdot \varepsilon, \\[4pt]  
   \Dgen \proj{S} (\Dgen \proj{J} 
         \varepsilon) 
      + \Dgen \proj{S} (\Dgen 
         \proj{S} \varepsilon) 
      &= \GenS\, \varepsilon,
\end{aligned}
\end{equation}
where $\GenS$ and $\GenRic_{AB}$ provide the scalar and non-scalar parts of $\GenRicci_{AB}$ respectively\footnote{Note
  that $\adj{P}^\perp \subset (S\otimes J) \oplus \bbR$ and
  the $\dHd$ structure gives an isomorphism $S\simeq S^*$ and
  $J\simeq J^*$. Thus, as in the first line of~\eqref{eq:GenRS}, we can
  also view $\GenRic$ as a map from $S$ to $J$.}. The existence of expressions of this type is a non-trivial statement. By computing in the split frame, it can be shown that the LHS is linear in $\varepsilon$, and since $\varepsilon$ and the LHS are manifestly covariant, these expressions define a tensor. We will write the components explicitly in section~\ref{sec:reform}, equation~\eqref{Rinframe}. This calculation further provides the non-trivial result that $\GenRicci_{AB}$ is restricted to be a section of $\adj{P}^\perp$, rather than a more general section of $(S \otimes J) \oplus \bbR$. In the context of supergravity, this calculation exactly corresponds to the closure of the supersymmetry algebra on the fermionic equations of motion, as will be discussed further in~\cite{CSW2}. Finally, since it is built from unique operators, the generalised curvature is automatically unique for a torsion-free compatible connection.

The expressions~\eqref{eq:GenRS} can be written with a different
sequence of projections. This helps elucidate the nature of the curvature
in terms of certain second-order differential operators. In
conventional differential geometry the commutator of two connections
$[\nabla_m, \nabla_n]$ has no second-derivative term simply because
the partial derivatives commute. This is a necessary condition for the
curvature to be tensorial. In $\Edd$ indices one can similarly write the
commutator of two generalised derivatives formally as   
%
	$(\Dgen \wedge \Dgen)_{AB} = [ \Dgen_A , \Dgen_B]$.
%
More precisely, acting on an $\Edd\times\bbR^+$ vector bundle $X$ we have 
\begin{equation}
	(\Dgen \wedge \Dgen): X \rightarrow \Lambda^2 E^* \otimes X. 
\end{equation}
Since again the partial derivatives commute this operator contains no
second-order derivative term, and so can potentially be used to
construct a curvature tensor. However, in $\Edd\times\bbR^+$
generalised geometry we also have $\der f \proj{N^*} \der g = 0$
for any $f$ and $g$, and so we can take the projection to the
bundle $N^*$ defined earlier, giving a similar operator  
\begin{equation}
	(\Dgen \proj{N^*} \Dgen) : X \rightarrow N^* \otimes X, 
\end{equation}
which will again contain no second-order derivatives. One thus expects
that these two operators, which can be defined for an arbitrary
$\Edd\times\bbR^+$ connection, should appear in any definition of
generalised curvature. Given an $\dHd$ structure and a
torsion-free compatible connection $D$, they indeed enter the definition of
$\GenRicci_{AB}$. Using $\dHd$ covariant projections one finds 
\begin{equation}
\label{eq:GenRS2}
\begin{aligned}
   (\Dgen \wedge \Dgen) \proj{J} \varepsilon 
      + (\Dgen \proj{N^*} \Dgen) \proj{J} \varepsilon 
      &= \GenRic \cdot \varepsilon , \\
   (\Dgen \wedge \Dgen) \proj{S} \varepsilon 
      + (\Dgen \proj{N^*} \Dgen) \proj{S} \varepsilon 
      &= \GenS \,\varepsilon . 
\end{aligned}
\end{equation}
This structure suggests there will be similar definitions of curvature
in terms of the operators $(\Dgen \wedge \Dgen)$ and $(\Dgen
\proj{N^*} \Dgen)$ independent of the representation on which
they act, and potentially without the need for additional structure.  


\section{Supergravity as $\Hd$ generalised geometry}
\label{sec:sugra-rel}

We now show how the generalised geometrical structures we have described in the previous sections allow us to rewrite the bosonic sector of eleven-dimensional supergravity with the local $\Hd$-covariance manifest. We also cover the relation to type II theories and the embedding tensor formalism.


\subsection{Eleven-dimensional supergravity in $d$-dimensions}
\label{sec:sugra}

We will be interested in ``restrictions'' of eleven-dimensional
supergravity where the spacetime is assumed to be a product
$\bbR^{10-d,1}\times M$ of Minkowski space with a $d$-dimensional
spin manifold $M$, with $d\leq 7$. The metric is taken to be  
\begin{equation}
   \dd s_{11}^2 = \ee^{2\Delta}\dd s^2(\bbR^{10-d,1}) + \dd s_d^2(M) , 
\end{equation}
where $\dd s^2(\bbR^{10-d,1})$ is the flat metric on $\bbR^{10-d,1}$ and $\dd
s_d^2(M)$ is a general metric on $M$. The warp factor $\Delta$ and all the
other fields are assumed to be independent of the flat $\bbR^{10-d,1}$
space. In this sense we restrict the full eleven-dimensional theory to
$M$. We will split the eleven-dimensional indices as external indices $ \mu = 0, 1, \dots , \dex-1$ and internal indices $m=1,\dots,d$ where
$\dex+d=11$. The full eleven-dimensional theory and the conventions we
are using are summarised in appendix~\ref{app:11d}. 

In the restricted theory, the surviving fields include the obvious
internal components of the eleven-dimensional fields (namely the
metric $g$ and three-form $A$) as well as the warp factor $\Delta$. If
$d=7$, the eleven-dimensional Hodge dual of the 4-form $F$ can have a
purely internal 7-form component. This leads 
one to introduce in addition a dual six-form potential $\tilde{A}$ on
$M$ which is related to the seven-form field strength $\tilde{F}$ by 
\begin{equation}
   \tilde{F} = \dd \tilde{A} - \tfrac12 A \wedge F ,
\end{equation}
The Bianchi identities satisfied by $F=\dd A$ and $\tilde{F}$ are then 
\begin{equation}
\begin{aligned}
   \dd F &= 0 , \\
   \dd \tilde{F} + \tfrac12 F \wedge F &= 0 . 
\end{aligned}
\end{equation}
With these definitions we see that $F$ and $\tF$ are related to the eleven dimensional 4-form field strength $\mathcal{F}$ by
\begin{equation}
	F_{m_1 \dots m_4} = \mathcal{F}_{m_1 \dots m_4} ,
		\qquad \qquad \qquad \tF_{m_1 \dots m_7} = \left(*_{\scriptscriptstyle 11}\mathcal{F}\right)_{m_1 \dots m_7}.
\end{equation}

One obtains the internal bosonic action
\begin{equation}
\label{eq:Boson-action}
   S_{\text{B}} = \frac{1}{2\kappa^2}\int \sqrt{g} \; \ee^{\dex \Delta} \left(
        \Scalar + \dex(\dex-1) (\der \Delta)^2
          - \tfrac12 \tfrac{1}{4!} F^2 - \tfrac12 \tfrac{1}{7!} \tilde{F}^2 \right) ,
\end{equation}
by requiring that its associated equations of motion 
\setlength{\jot}{8pt}
\begin{equation}
\label{eq:eom11d}
\begin{aligned}
   \Ric_{mn} - \dex \LC_m \LC_n \Delta - \dex(\der_m \Delta)(\der_n \Delta) 
       - \tfrac12 \tfrac{1}{4!}\Big( 
             4F_{m p_1 p_2 p_3} F_n{}^{p_1 p_2 p_3}
             -& \tfrac{1}{3} g_{mn} F^2 \Big) \\
       - \tfrac12 \tfrac{1}{7!}\left(  7\tilde{F}_{m p_1 \dots p_6} 
                \tilde{F}_n{}^{p_1 \dots p_6}
             - \tfrac{2}{3} g_{mn} \tilde{F}^2 \right) &= 0 , \\
	\Scalar - 2(\dex-1) \LC^2 \Delta - \dex(\dex-1) (\der \Delta)^2 
                - \tfrac12 \tfrac{1}{4!} F^2 
		- \tfrac12 \tfrac{1}{7!} \tilde{F}^2 
                &= 0 , \\
	\dd * (\ee^{\dex \Delta}F) 
            - \ee^{\dex \Delta} (*\tilde{F})\wedge F &= 0, \\
	\dd * (\ee^{\dex \Delta} \tilde{F}) &= 0. 
\end{aligned}
\end{equation}
are those obtained by substituting the field ansatz into~\eqref{eq:eom11}. 

Although here we are interested in the bosonic sector of supergravity,
note that the supersymmetry variations of the gravitino can also be
written as  
\begin{equation}
\label{eq:susy}
\begin{aligned}
   \delta \rho &= 
      \gamma^m \LC_m \epsilon 
      - \tfrac{1}{4} \tfrac{1}{4!} \gamma^{m_1 \dots m_4}F_{m_1 \dots m_4} \epsilon
         - \tfrac{1}{4} \tfrac{1}{7!} \tilde{F}_{m_1 \dots m_7} \gamma^{m_1 \dots m_7} \epsilon + \tfrac{\dex-2}{2}
         (\gamma^m\der_m \Delta)\epsilon ,\\    
   \delta \psi_m &= \LC_m \epsilon  
       + \tfrac{1}{288} \left(\gamma_m{}^{n_1 \dots n_4} 
          - 8 \delta_{m}{}^{n_1} \gamma^{n_2 n_3 n_4} \right) 
		F_{n_1 \dots n_4} \epsilon
          - \tfrac{1}{12} \tfrac{1}{6!} \tilde{F}_{mn_1 \dots n_6} 
			\gamma^{n_1 \dots n_6} \epsilon ,
\end{aligned}
\end{equation}
where $\rho$ is related to the trace of the gravitino in the external
space and $\gamma^m$ are $\Cliff(d;\bbR)$ gamma matrices. These
expressions will be discussed in more detail in~\cite{CSW2}.


\subsection{Reformulation as $\Hd$ generalised geometry}
\label{sec:reform}

It is well known~\cite{cremmer,julia} that the bosonic fields of
the reduced supergravity parametrise a $(\Edd\times\bbR^+)/H_d$ coset,
that is, at each point $x\in M$,  
\begin{equation}
\label{eq:coset-fields}
   \{ g, A, \tA, \Delta \} 
      \in \frac{\Edd}{\Hd}\times\bbR^+ . 
\end{equation}
Thus giving the bosonic fields is equivalent to specifying a
generalised metric $G$. Furthermore, the infinitesimal bosonic symmetry
transformation (diffeomorphisms and gauge transformations of $A$ and
$\tA$) are encoded by the Dorfman derivative~\cite{GMPW}
\begin{equation}
   \delta_V G = \Lgen_V G ,  
\end{equation}
and the algebra of these transformations is given by the Courant
bracket. 

We now show that the dynamics of the reduced theory are encoded by the
torsion-free $\Hd$ connection $\Dgen$. By doing so we show that the
theory can be reformulated geometrically with a local $\Hd$
invariance. Such local symmetries were first considered, for $d=7$, by
de Wit and Nicolai~\cite{deWN}. The generalised geometry here can be
viewed as a geometrical explanation of their original rewriting. In
order to match the dynamics we work in a particular frame, namely 
the conformal split $\Hd$ frame, which is equivalent to an $O(d)$
structure on $E$. It is worth stressing that the generalised
geometrical theory is $\Hd$ covariant, it is simply that supergravity
is conventionally written with only an $O(d)\subset\Hd$ manifest. 

We have already seen that in the conformal split frame $\Dgen$ takes
the form~\eqref{eq:Dgen-sol}. Viewing sections of $S$ and $J$ in
$\Spin(d)$ representations one can then write the unique
operators~\eqref{eq:unique-ops} in this basis. For example, taking $\epsilon = \ee^{\Delta/2} \varepsilon$ to be the supersymmetry parameter, one finds
\begin{equation}
\label{eq:D-SUSY}
\begin{aligned}
   \ee^{-\Delta/2} (\Dgen \proj{J} \varepsilon )_a 
      &= \LC_a \epsilon 
          + \tfrac{1}{288} \left(\gamma_a{}^{b_1 \dots b_4} 
             - 8 \delta_{a}{}^{b_1} \gamma^{b_2 b_3 b_4} \right) 
		F_{b_1 \dots b_4} \epsilon
          \\ & \qquad \qquad \qquad 
          - \tfrac{1}{12} \tfrac{1}{6!} \tilde{F}_{ab_1 \dots b_6} 
			\gamma^{b_1 \dots b_6} \epsilon , \\
  \ee^{-\Delta/2} (\Dgen \proj{S} \varepsilon ) 
      &= \gamma^m \LC_m \epsilon 
      	- \tfrac{1}{4} \tfrac{1}{4!} \gamma^{m_1 \dots m_4}F_{m_1 \dots m_4} \epsilon
        \\ & \qquad \qquad \qquad  
        	- \tfrac{1}{4} \tfrac{1}{7!} \tilde{F}_{m_1 \dots m_7} \gamma^{m_1 \dots m_7} \epsilon 
        + \tfrac{\dex-2}{2} (\gamma^m\der_m \Delta)\epsilon .
\end{aligned}
\end{equation}
These exactly match the operators that appear in the supersymmetry
variations~\eqref{eq:susy}. Given such expressions one can then
calculate the Ricci tensor~\eqref{eq:GenRS} in this frame, finding, 
\begin{equation}
\label{Rinframe}
\begin{aligned}
   \ee^{-2\Delta} \GenRicci_{ab} &=
      \Ric_{ab} - \dex \LC_a \LC_b \Delta 
         - \dex(\der_a \Delta)(\der_b \Delta) 
         \\ & \qquad 
         - \tfrac12 \tfrac{1}{4!}\Big( 
            4F_{a c_1 c_2 c_3} F_b{}^{c_1 c_2 c_3}
            - \tfrac{1}{3} g_{ab} F^2 \Big) 
         \\ & \qquad 
       - \tfrac12 \tfrac{1}{7!}\left(  7\tilde{F}_{a c_1 \dots c_6} 
                \tilde{F}_b{}^{c_1 \dots c_6}
             - \tfrac{2}{3} g_{ab} \tilde{F}^2 \right) , \\
    \ee^{-2\Delta} \GenRicci_{abc} &= 
      \tfrac12 \left[ \ee^{-\dex \Delta}* \dd * \ee^{\dex \Delta}F 
         -  * (* \tilde{F} \wedge F)
         \right]_{abc} ,\\ 
    \ee^{-2\Delta} \GenRicci_{a_1 \dots a_6} &= \tfrac12 \left[
      \ee^{-\dex \Delta} * \dd * \ee^{\dex \Delta} \tilde{F} \right]_{a_1 \dots a_6},\\
    \ee^{-2\Delta} \GenS &= \Scalar - 2(\dex-1) \LC^2 \Delta - \dex(\dex-1) (\der \Delta)^2 
                - \tfrac12 \tfrac{1}{4!} F^2 
		- \tfrac12 \tfrac{1}{7!} \tilde{F}^2 .
\end{aligned}
\end{equation}

Comparing with~\eqref{eq:Boson-action} and~\eqref{eq:eom11d} we see
that the bosonic action is given by 
\begin{equation}
\label{eq:S-gen}
   S_{\text{B}} = \int \vol_G \GenS , 
\end{equation}
where $\vol_G$ is the $\Edd$-invariant scalar given
in~\eqref{eq:Phi-def}, and that the bosonic equations of motion are
equivalent to  
\begin{equation}
\label{eq:eom-gen}
   \GenRicci_{MN} = 0 .
\end{equation}
As advertised, we have rewritten the bosonic dynamics in terms of
generalised curvatures with a manifest $\Hd$ local symmetry.


\subsection{Relation to type II supergravity}
\label{sec:typeII}

The $(\Edd\times\bbR^+)/\Hd$ coset structure can equally well describe
the fields of type II theories in $d-1$ dimensions. Specifically 
\begin{equation}
\label{eq:coset-fields-tII}
   \{ g, B, \tilde{B}, \phi, A^\pm, \Delta \} 
      \in \frac{\Edd}{\Hd}\times\bbR^+ , 
\end{equation}
where $B$ is the NSNS two-form field, $\tilde{B}$ is the six-form
potential dual to $B$, $\phi$ is the dilaton and
$A^\pm$ are the RR potentials (in a democratic formalism) where $A^-$
is a sum of odd-degree forms in type IIA and $A^+$ is a sum of
even-degree forms in type IIB. All the fields now depend on a $d-1$
dimensional manifold $M'$. 

The construction of torsion-free compatible connections $\Dgen$
follows exactly as above. The only difference is that the bundles,
partial derivative and split frames are now naturally written in terms
of a $\GL(d-1,\bbR)$ subgroup of $\Edd$ as opposed to
$\GL(d,\bbR)$ (the appropriate subgroups are defined in
appendix~\ref{app:typeII}.) In particular, the generalised tangent space
takes the form~\cite{chris,E7-flux,GLSW,GO}
\begin{equation}
\label{eq:EisoII}
   E \simeq TM' \oplus T^*M' \oplus \Lambda^5T^*M' 
         \oplus (T^*M' \otimes\Lambda^6T^*M')
         \oplus \Leo T^*M' ,
\end{equation}
where ``even'' refers to type IIA and ``odd'' to IIB. The partial
derivative $\der$ now acts via the embedding $T^*M'\to E^*$ so that 
$\der f=\dd f\in\Gs{E^*}$, for any function $f$. One still has the
``section condition'' $\der f \proj{N^*}\der g=0$ but now the space
spanned by $\der f$ is not the maximal such subspace in $E^*$ (since
$\der_M$ spans one less dimension). 

We will not give the expressions for the type II decompositions here,
though given the $\Spin(d-1)$ spinor decomposition in
appendix~\ref{app:typeII}, it is relatively straightforward to
calculate them directly from $\Spin(d)$ expressions given in the
previous section. The central point is that the bosonic equations of
motion and action given by~\eqref{eq:S-gen} and~\eqref{eq:eom-gen} are left unchanged. What changes is the decomposition of these
expressions in the bosonic fields, and the partial derivative action
$\der$. 


\subsection{Identity structures, fluxes and relation to the embedding
  tensor}  
\label{sec:embedT}

The embedding tensor is the object that determines general gaugings
of (typically maximally) supersymmetric theories in $11-d$
dimensions~\cite{embedT1,embedT2}. It is striking that it transforms in the
same $\Edd$ representations as the generalised
torsion~\eqref{eq:Tgen-def}, namely $K\oplus E^*$. That such
representations appear in gauged supersymmetric theories has been
discussed in detail in~\cite{E11-gauged} in the context of $E_{11}$
theory (as well as~\cite{e10-embed} in the case of
$E_{10}$). Here we simply show why, in the current context,
the generalised torsion and the embedding tensor are related when the 
gauged supergravity arises from a dimensional reduction of
eleven-dimensional supergravity on a $d$-dimensional manifold
$M$. This also relates to the observation that the generic
set of fluxes, both geometrical and non-geometrical, are sections
of the same bundle $K$~\cite{E7-flux}.

To make the connection we first need to identify what structures on
the internal space $M$ lead to maximally supersymmetric theories in
$11-d$ dimensions. Metrically the eleven-dimensional space is a
fibration
\begin{equation}
   \dd s^2 = \hat{g}_{\mu\nu}(y) \dd y^\mu\dd y^\nu
                + g_{mn}(x,y)(\dd x^m + A^m(x,y))(\dd x^n + A^n(x,y)),
\end{equation}
where $y$ and $x$ are coordinates on the external and internal spaces
respectively, and $A^m$ are one-forms on the external space. It is
well known that tori give suitable supersymmetric compactifications, as
do generic twisted tori (or local group manifolds)~\cite{twisted},
including their non-geometrical
extensions~\cite{non-geom-twisted,dft-ss,GMPW}. The corresponding
relation  to the embedding tensor is also well established (for a review
see~\cite{flux-gauge-review}). The characteristic feature of these
backgrounds is that the associated generalised 
tangent space $E$ admits an ``identity structure'', that is, a
$G$-structure $P\subset\tilde{F}$ where $G$ is just the trivial group,
with a single element, the identity. This means that the space admits a
single \emph{globally defined} frame $\{\hat{E}^A\}$ and is a
necessary condition for a reduction to a maximally supersymmetric
effective theory. As discussed in the context of $N=2$ supersymmetry
in~\cite{GLW,GLSW}, such reductions require globally defined spinors on 
$M$. For a maximally symmetric theory, there is a maximal number of
such spinors and hence the $\dHd$-bundle is trivial, implying we have
an identity structure 
\begin{equation}
   \left\{
      \pbox{\textwidth}{maximal supersymmetric\\ effective theory}
   \right\}
   \Longleftrightarrow
   \left\{
      \pbox{\textwidth}{exists globally defined\\ 
         frame on $M$}
   \right\} . 
\end{equation}
Such structures are also sometimes referred to as
``generalised parallelizable'' spaces~\cite{GMPW}. They are the
generalised analogues of parallelizable spaces, where there is a
globally defined frame $\{\hat{e}^a\}$. Note that twisted tori give
examples of such generalised parallelizable spaces, but in principle
one could also have a generalised parallelization of $E$ without a
parallelization of $TM$. 

In making the connection to the embedding tensor let us focus on the
scalar moduli fields, which parametrise the coset $\Edd/\Hd$. Recall
that given a conventional global frame $e^a(x)$ (for example
left-invariant one-forms on a local group manifold) one can write a
family of frames $e^{\prime a}(x,y)=m^a{}_b(y) e^b(x)$ and make a
Scherk--Schwarz reduction~\cite{ss}. The corresponding
metrics are given by 
\begin{equation}
   g' = \delta_{ab}\, e^{\prime a}(x,y)\otimes e^{\prime b}(x,y) 
         = h_{ab}(y) \, e^a(x)\otimes e^b(x) ,
\end{equation}
where $h_{ab}(y)=\delta_{cd}m^c{}_a(y)m^b{}_d(y)$ are moduli
parametrising $\GL(d,\bbR)/O(d)$. Now suppose we have a generalised
parallelization $\hat{E}_A(x)$ and a dual basis $E^A(x)$. The scalar
fields in the effective theory similarly can be regarded as
parametrising generic $\Edd$ transformations 
$E^{\prime A}(x,y)=M^A{}_B(y)E^B(x)$, defining the generalised metric,  
\begin{equation}
\label{eq:G-id}
   G' = \delta_{AB}\, E^{\prime A}(x,y)\otimes E^{\prime B}(x,y) 
         = H_{AB}(y) \, E^A(x)\otimes E^B(x) ,
\end{equation}
where $H_{AB}(y)=\delta_{CD}M^C{}_A(y)M^B{}_D(y)$ are moduli
parametrising an $\Edd/\Hd$ coset. Note that we ignore the $\bbR^+$ degree
of freedom that rescales $G$. Since this factor was associated with
warping of the external space, which in the dimensionally reduced
theory is encoded in the conformal rescaling of the external metric
$\hat{g}$, this does not lose any degrees of freedom. 

The potential for the scalar moduli arises from the dimensional
reduction of the action on the internal space, which, as we have seen,
can be written in terms of the generalised Ricci scalar as
in~\eqref{eq:S-gen}. This in turn is completely determined by the
torsion-free connection $G'$-compatible connection $D'$. One can
construct $D'$ as follows. Given the transformed frame
$\{\hat{E}'_A\}$ we can always define a connection $\Dgen''$ that is
compatible with the corresponding identity structure, that is, for all
$A$,   
\begin{equation}
   \Dgen'' \hat{E}'_A = 0, 
\end{equation}
but in general it will be torsionful. By definition the torsion is
simply given by the algebra of the basis $\{\hat{E}'_A\}$ under the Dorfman
derivative, namely,
\begin{equation}
\label{eq:T-id}
   \Lgen_{\hat{E}'_A} \hat{E}'_C 
      = - T'_A{}^B{}_C \hat{E}'_B ,
\end{equation}
where $T'\in\Gs{K\oplus E^*}$ and is moduli dependent. By construction
$\Dgen''$ is compatible with the generalised
metric~\eqref{eq:G-id}. The torsion-free metric compatible connection
can then be constructed as  
\begin{equation}
   D' = D'' + \Sigma' , 
\end{equation}
where, in the notation of section~\ref{sec:genLC},
$\Sigma'\in\Gs{E^*\otimes\adj{P}}$ and for $\Dgen'$ to be torsion-free
we require 
\begin{equation}
   \Sigma' = - T' + \am' , 
\end{equation}
where the ambiguous part $\am'\in\Gs{U}$. Since the supergravity does
not depend on the ambiguous part $\am'$ we see that effective theory,
and in particularly the scalar potential, is determined completely by the
moduli-dependent $T'$ defined in~\eqref{eq:T-id}.

We could also consider the corresponding tensor $T$ for the fixed frame
$\{\hat{E}_A\}$. This is independent of the moduli, and is related to $T'$, the
corresponding ``dressed'' version, simply by transforming indices
with $M$ or $M^{-1}$ as appropriate. The undressed $T$ can be directly
identified with the embedding tensor if we make the further assumption
that its components $T^{\ph{A}B}_{A\ph{B}C}$ are constant. First we
note that it is in the same representations of $\Edd$ as the embedding  
tensor. Second it satisfies the same quadratic
relation~\cite{embedT1}, giving the embedding of the gauged symmetry 
group in $\Edd$. Here this condition arises from the Jacobi-like
relation, following from the fact that $\Lgen_U$ satisfies the Leibniz
identity,  
\begin{equation}
   \Lgen_U(\Lgen_V W) - \Lgen_V(\Lgen_U W) = \Lgen_{\Lgen_U V}W . 
\end{equation}
Taking $U=\hat{E}^A$, $V=\hat{E}^B$ and $W=\hat{E}^C$ this gives
\begin{equation}
   \BLie{T_A}{T_B} = T_{A}{}^C{}_{B} T_C , 
\end{equation}
where we view $(T_A)^B{}_C=T_A{}^B{}_C$ as a set of elements in the
adjoint representation of $\Edd$ labelled by $A$.

We can then make the connection to~\cite{E7-flux}, where it was shown
that generic fluxes in string compactifications correlate with
the embedding tensor. The definition~\eqref{eq:T-id} gives $T$ a
direct geometrical interpretation which matches the fluxes identified
in~\cite{E7-flux} in the context of type IIB backgrounds. Suppose, for
example we have a twisted torus (that is a local group manifold) with
a global frame $\hat{e}_a$. We can then define a generalised
parallelization by taking $\{\hat{E}^A\}$ in the split
form~\eqref{eq:geom-basis} (that is, with $\Delta=0$). Let $\nabla$ be 
the conventional connection that satisfies $\nabla \hat{e}_a=0$ and
has torsion $T^a{}_{bc}=-f_{bc}{}^a$ where $f$ are the structure
constants given by
$\BLie{\hat{e}_a}{\hat{e}_b}=f_a{}^c{}_b\hat{e}_c$. By definition we
then have that $\Dgen^\nabla \hat{E}_A=0$. Using~\eqref{eq:splitT}, we can 
calculate the components of $T$ of the torsion of
$\Dgen^\nabla$ and find 
\begin{equation}
   T(V) = - i_v f - i_v F - i_v\tF + \omega\wedge F , 
\end{equation}
where $(i_v f)^a{}_b=v^c f_{cb}{}^a$ is a section of $TM\otimes
T^*M$. Thus only certain components of $T$, the so-called geometrical
fluxes, are non-zero. The corresponding split frame for type IIB
generates the geometrical fluxes identified in~\cite{E7-flux}.


\section{Conclusions and discussion}
\label{sec:conc}


We have seen that the action, equations of motion and symmetries for
the bosonic fields in reductions of M theory to $d$ dimensions
actually have a remarkably simple and unified form. The fields unify
as a generalised metric. The symmetries are simply the generalised
geometry extensions of diffeomorphisms, and the action is simply the
analogue of the Ricci scalar. The formalism works for all dimensions
$d\leq 7$ and the theory has an extended local $\Hd\supset O(d)$
symmetry. It is a direct extension of our earlier work~\cite{CSW1} on
reformulating type II supergravities using $O(10,10)\times\bbR^+$
generalised geometry. 

It is natural to ask how this structure might extend to higher
dimensions. The basic obstruction, even for $d=8$, is that although the
generalised tangent space exists, together with an
$E_{8(8)}\times\bbR^+$ structure bundle, and a Dorfman derivative, one
cannot write the derivative in the form~\eqref{eq:Lgen-cov}, since
this gives a non-covariant expression. Consequently, one does not have
a natural way to define the generalised torsion. The problem with
writing the derivative in this form is the presence of the 
$\tau\in\Gs{T^*M\otimes\Lambda^7T^*M}$ tensors in $E$. Physically
these correspond to Kaluza--Klein monopole charges in the U-duality
algebra and should be associated to the symmetries of ``dual
gravitons''. Note that these terms already meant, even in $d=7$, that
we could not write the difference of between the Dorfman derivative
and the bracket~\eqref{eq:bracket-diff} an the $\Edd\times\bbR^+$
covariant form. One possible way out is to include dependence on the
``non-compact'' $(11-d)$-dimensional space $M_{\text{ext}}^{10-d,1}$. Allowing for
diffeomorphisms in $M_{\text{ext}}^{10-d,1}$ may then correct the non-covariance of the
naive structure. 

The possibility of extending the formalism to the Kac--Moody algebras
$E_{10}$ or $E_{11}$ is particularly intriguing. Since we assume an
underlying manifold, the connection to the $E_{10}$ formalism
of~\cite{e10} is less direct, since there the spacetime is emergent,
the $E_{10}$ fields encoding a spatial gradient expansion around a
spacetime point. The $E_{11}$ formalism on the other hand  starts with
(an infinite number) of coordinates~\cite{west-coord} transforming in
a particular representation $l_1$ (which corresponds to the
generalised tangent space representation upon reduction to
$\Edd$). One important question is how the dependence on these
coordinates might be truncated to define eleven-dimensional
supergravity. The results here would suggest one imposes a quadratic
section condition~\eqref{eq:section-cond} projecting onto
the corresponding $N^*$ representation defined by the appropriate node
in the $e_{11}$ Dynkin diagram as described in
section~\ref{sec:gen-tensor}. Another very interesting possibility is
that, if a generalised geometrical formulation can be found for $d>8$
it may be that the existence of the torsion-free compatible connection
$\Dgen$ actually constrains the $\Hd$-structure. This is what happens
for instance with conventional connections compatible with an almost
complex structure, where the existence of a torsion-free compatible
connections requires the structure to be integrable. Such a situation
would impose differential conditions on the fields defining the coset
$(\Edd\times\bbR^+)/\Hd$, perhaps truncating the infinite set to a
finite number of independent components corresponding to the degrees
of freedom of supergravity. This may be connected to the recent
result~\cite{HKN} that, given fairly weak assumptions, only a finite
number of the fields in the Kac--Moody algebra are propagating. 

As we have already stressed, the formalism used here and
in~\cite{CSW1} is locally equivalent to the standard formulation of
double field theory and its M theory variants. The
derivations relied only on the partial derivative satisfying the
section condition~\eqref{eq:section-cond} (or the corresponding
condition for $O(d,d)$). The maximal subspace of $E^*$ satisfying this
condition is $d$-dimensional and is stabilised by the parabolic
subgroup $\Ggeom$. In the context of double field theory it defines
the set of coordinates on which the fields depend, and is a necessary
condition for the formulation of an action and a closed symmetry
algebra. This defines a foliation and reducing along the isometries, the theory is locally defined on
a $d$-dimensional manifold as in generalised geometry. In either
formulation there is a global $O(d,d)$ or $\Edd$ symmetry acting on
the frame bundle. However, while the strong constraint is covariant,
the particular choice of a maximal subspace is not invariant under the
global symmetry group. 

There are number of other natural extensions to the geometry described
here.  It would be interesting to understand if similar
constructions can be used for other supergravity theories.
One might also wonder if the formalism can be used to describe
higher-derivative corrections and their $\Edd$ transformation
properties. A more direct, key application is the description of supersymmetric
backgrounds. Formulations of $N=1$ and $N=2$ backgrounds in
$E_{7(7)}$ language have already been given in~\cite{GO}. In the
current context one expects that generic supersymmetric backgrounds in
$d\leq 7$ should correspond to special holonomy $G\subset\dHd$ for the
operator $\Dgen$. Note that it is the holonomy of $\Dgen$ and not its
projections~\eqref{eq:D-SUSY} that are relevant, and hence $G$ is
a subgroup of $\dHd$. This is in contrast to~\cite{gen-holo} where the
holonomy of the operators appearing directly in the
supersymmetry variations was considered, and larger groups can
appear. The most obvious extension, though, is the completion of the description of the supergravity by
including the fermionic degrees of freedom and supersymmetry
transformations, at least to leading order. This will be the main result of~\cite{CSW2}. 


\acknowledgments

We would like to thank Mariana Gra\~{n}a, Chris Hull and Barton
Zwiebach for helpful discussions. This work was supported by the STFC
grant ST/J000353/1. C.~S-C.~is supported by an STFC PhD
studentship. A.~C.~is supported by the Portuguese Funda\c c\~ao para a
Ci\^encia e a Tecnologia under grant SFRH/BD/43249/2008. D.~W.~also
thanks CEA/Saclay and the Mitchell Institute for Fundamental Physics
and Astronomy at Texas A\&M for hospitality during the completion of
this work.


\appendix


\section{Eleven-dimensional supergravity}
\label{app:11d}

Let us start by summarising the action, equations of motion and
supersymmetry variations of eleven-dimensional supergravity, to
leading order in the fermions, following the conventions
of~\cite{GP}. 

The fields are simply 
\begin{equation}
   \{ g_{\mu\nu}, \mathcal{A}_{\mu\nu\rho}, \psi_\mu\} , 
\end{equation}
where $g_{\mu\nu}$ is the metric, $\mathcal{A}_{\mu\nu\rho}$ the three-form
potential and $\psi_\mu$ is the gravitino. The bosonic action is given by

\begin{equation}
\label{eq:NSaction}
   S_{\text{B}} = \frac{1}{2\kappa^2}\int \left(
        \vol_g\Scalar - \tfrac{1}{2}\mathcal{F}\wedge *\mathcal{F}
          - \tfrac{1}{6}\mathcal{A}\wedge \mathcal{F}\wedge \mathcal{F} \right) ,
\end{equation}
where $\Scalar$ is the Ricci scalar and $\mathcal{F}=\dd \mathcal{A}$. This leads to the equations of motion 
\begin{equation}
\label{eq:eom11}
\begin{aligned}
   \Ric_{\mu\nu} - \tfrac{1}{12} \left( 
           \mathcal{F}_{\mu\rho_1\rho_2\rho_3} \mathcal{F}^{\ph{\nu}\rho_1\rho_2\rho_3}_\nu
		- \tfrac{1}{12} g_{\mu\nu} \mathcal{F}^2 \right)
      &= 0 , \\
   \dd * \mathcal{F} + \tfrac{1}{2} \mathcal{F}\wedge \mathcal{F} &= 0 , 
\end{aligned}
\end{equation}
where $\Ric_{\mu\nu}$ is the Ricci tensor. Note that we are using the
notation $K^2=K_{\mu_1\dots\mu_k}K^{\mu_1\dots\mu_k}$ for a rank
$k$ tensor $K$. 

The supersymmetry variation of the gravitino is
\begin{equation}
   \delta \psi_\mu 
      = \LC_\mu \epsilon + \tfrac{1}{288} \left(
         \Gamma_\mu{}^{\nu_1 \dots \nu_4}
            - 8 \delta_\mu{}^{\nu_1} \Gamma^{\nu_2 \nu_3 \nu_4} \right) 
         \mathcal{F}_{\nu_1 \dots \nu_4} \epsilon ,
\end{equation}
where $\Gamma^\mu$ are the $\Cliff(10,1;\bbR)$ gamma matrices and $\epsilon$ is the supersymmetry parameter.


\section{Conventions in Euclidean signature}
\label{app:conv-d}

We use the indices $m,n,p, \dots $ as the coordinate indices and
$a,b,c \dots$ for frame indices. We take symmetrisation of
indices with weight one. Given a polyvector $w\in\Lambda^pTM$ and a
form $\lambda\in\Lambda^qT^*M$, we write in components 
\begin{equation}
\begin{aligned}
   w &= \frac{1}{p!} w^{m_1 \dots m_p} 
      \frac{\der}{\der x^{m_1}} \wedge \dots 
          \wedge \frac{\der}{\der x^{m_p}} , \\ 
   \lambda &= \frac{1}{q!} \lambda_{m_1 \dots m_q} 
       \dd x^{m_1} \wedge \dots \wedge \dd x^{m_q} , 
\end{aligned}
\end{equation}
so that wedge products and contractions are given by 
\begin{equation}
\begin{aligned}
   \left(w \wedge w'\right)^{m_1\dots m_{p+p'}} 
       &= \frac{(p+p')!}{p!p'!}
          w^{[m_1 \dots m_p}w'^{m_{p+1} \dots m_{p+p'}]} , \\
   \left(\lambda \wedge \lambda'\right)_{m_1\dots m_{q+q'}} 
       &= \frac{(q+q')!}{q!q'!}
          \lambda_{[m_1 \dots m_q} \lambda'_{m_{q+1} \dots m_{q+q'}]} , \\
   \left(w\inn \lambda\right)_{m_1\dots m_{q-p}} 
      &:= \frac{1}{p!} w^{n_1\dots n_p}
         \lambda_{n_1\dots n_p m_1\dots m_{q-p}} &&&& 
         \text{if $p\leq q$} , \\
   \left(w\inn \lambda\right)^{m_1\dots m_{p-q}} 
      &:= \frac{1}{q!} w^{m_1\dots m_{p-q}n_1\dots n_q}
         \lambda_{n_1\dots n_q} &&&& 
         \text{if $p \geq q$} .  
\end{aligned}
\end{equation}

Given the tensors $t\in TM\otimes\Lambda^7TM$, $\tau\in
T^*M\otimes\Lambda^7T^*M$ and $a\in TM\otimes T^*M$ with components 
\begin{equation}
\begin{aligned}
   t &= \frac{1}{7!} w^{m,m_1 \dots m_7} 
      \frac{\der}{\der x^m} \otimes
          \frac{\der}{\der x^{m_1}}\wedge \dots 
          \wedge \frac{\der}{\der x^{m_7}} , \\ 
   \tau &= \frac{1}{7!} \tau_{m,m_1 \dots m_7} 
       \dd x^m \otimes \dd x^{m_1} \wedge \dots \wedge \dd x^{m_q} , \\
   a &= a^m{}_n \frac{\der}{\der x^m}\otimes \dd x^n , 
\end{aligned}
\end{equation}
and also a form $\sigma \in \Lambda^5 T^*M$ and a vector $v \in TM$, we use the ``$j$-notation'' from~\cite{PW}, defining 
\begin{equation}
\label{eq:jdef}
\begin{aligned}
   \left(w\inn \tau\right)_{m_1\dots m_{8-p}}
      &:= \frac{1}{(p-1)!} w^{n_1\dots n_p} 
         \tau_{n_1,n_2\dots n_p m_1\dots m_{8-p}} , \\
   \left(t\inn\lambda\right)^{m_1\dots m_{8-q}}
      &:= \frac{1}{(q-1)!} t^{n_1,n_2\dots n_q m_1\dots m_{8-q}}
         \lambda_{n_1\dots n_q} , \\
   \left(t\inn \tau\right) 
      & := \frac{1}{7!} t^{m,n_1\dots n_7}
         \tau_{m,n_1\dots n_7} , \\
   \left(jw\wedge w'\right)^{m,m_1\dots m_7}
      &:= \frac{7!}{(p-1)!(8-p)!}
         w^{m[m_1\dots m_{p-1}}w^{\prime m_p\dots m_7]} , \\
   \left(j\lambda\wedge\lambda'\right)_{m,m_1\dots m_7}
      &:= \frac{7!}{(q-1)!(8-q)!}
         \lambda_{m[m_1\dots m_{q-1}}\lambda'_{m_q\dots m_7]} , \\
   \left(jw\inn j\lambda\right)^m{}_n
      &:= \frac{1}{(p-1)!} w^{mn_1\dots n_{p-1}}
         \lambda_{nn_1\dots n_{p-1}} ,\\
   \left(jt\inn j\tau\right)^m{}_n
      &:= \frac{1}{7!} t^{m,n_1\dots n_7}
         \tau_{n,n_1\dots n_7} , \\ 
    \left( j^{p+1} \lambda \wedge \tau \right)_{m_1\dots m_{p+1}, n_1 \dots n_7} &:=
   (p+1) \lambda_{[m_1 \dots}\tau_{m_{p+1}],n_1 \dots n_7} ,\\
   \left( j^3 \sigma \wedge \sigma' \right)_{m_1\dots m_3, n_1 \dots n_7} &:=
   \tfrac{7!}{5! \cdot 2!} \sigma_{m_1 \dots m_3 [n_1 n_2}\sigma'_{\dots n_7]} ,\\ 
   (v \inn j\tau)_{mn_1 \dots n_6} &:= v^n \tau_{m,n n_1 \dots n_6} .
\end{aligned}
\end{equation}

The $d$-dimensional metric $g$ is always positive definite. We define
the orientation, $\epsilon_{1 \dots d} = \epsilon^{1 \dots d} = +1$,
and use the conventions
\begin{equation}
\label{eq:*norm}
\begin{aligned}
   * \lambda_{m_1\dots m_{d-q}} 
      &= \tfrac{1}{q!} \sqrt{|g|} 
         \epsilon_{m_1 \dots m_{d-q} n_1 \dots n_q} 
         \lambda^{n_1 \dots n_q} , \\
   \lambda^2 &= \lambda_{m_1\dots m_q}\lambda^{m_1\dots m_q} . 
\end{aligned}
\end{equation}
%


\section{$\Edd\times\bbR^+$ and $\Hd$} 
\label{app:Edd-embed}


\subsection{Construction of $\Edd\times\bbR^+$ from $\GL(d,\bbR)$}
\label{app:Edd-def}

In this section we give an explicit construction of $\Edd\times\bbR^+$
for $d\leq 7$ based on the $\GL(d,\bbR)$ subgroup.  
If $\GL(d,\bbR)$ acts linearly on the
$d$-dimensional vector space $F$, we define
\begin{equation}
\label{eq:Ed-bundles}
\begin{aligned}
      W_1 &=  F \oplus \Lambda^2F^* \oplus \Lambda^5F^*
         \oplus \left(F^*\otimes \Lambda^7F^* \right), \\
   W^*_1 &=  F^* \oplus \Lambda^2F \oplus \Lambda^5F
         \oplus \left(F\otimes \Lambda^7F \right) , \\
    W_{\text{ad}} &= \bbR \oplus \left(F\otimes F^*\right) 
        \oplus \Lambda^3F^* \oplus \Lambda^6F^* 
        \oplus \Lambda^3F \oplus \Lambda^6F  . 
\end{aligned}
\end{equation}
The corresponding $\Edd\times\bbR^+$ representations are listed in
Table~\ref{tab:gen-tang}. We write elements as
\begin{equation}
\label{eq:Ed-secs}
\begin{aligned}
   V &= v + \omega + \sigma + \tau && \in W_1 , \\
   Z &= \zeta + u + s + t && \in W_1^* , \\
   R &= c + r + a + \ta + \alpha + \talpha && \in W_{\text{ad}} ,
\end{aligned}
\end{equation}
so that $v\in F$, $\omega\in\ \Lambda^2F^*$, $\zeta\in F^*$,
$c\in\bbR$ etc. If $\{\hat{e}_a\}$ is a basis for $F$ with a dual
basis $\{e^a\}$ on $F^*$ then there is a natural $\gl(d,\bbR)$ action
on each tensor component. For instance 
\begin{equation}
\begin{aligned}
   (r\cdot v)^a &= r^a{}_b v^b , &&&
   (r\cdot \omega)_{ab} 
       &= - r^c{}_a \omega_{cb} - r^c{}_b \omega_{ac} , &&&
   \text{etc}. 
\end{aligned}
\end{equation}

Writing $V'=R\cdot V$ for the adjoint $\Edd\times\bbR^+$ action of
$R\in W_{\text{ad}}$ on $V\in F$, the components of $V'$, using the
notation of appendix~\ref{app:conv-d}, are given by 
\begin{equation}
\label{eq:E-tranfs}
\begin{aligned}
   v' &= c v + r\cdot v + \alpha\inn\omega - \talpha\inn \sigma  , \\
   \omega' &= c \omega + r\cdot \omega + v\inn a 
         + \alpha\inn \sigma + \talpha\inn\tau , \\
   \sigma' &= c \sigma + r\cdot \sigma + v\inn\ta
         + a\wedge \omega + \alpha\inn \tau , \\
   \tau' &= c \tau + r\cdot \tau 
         - j\ta \wedge \omega + ja\wedge \sigma . 
\end{aligned}
\end{equation}
Note that, the $\Edd$ sub-algebra is generated by setting $c=
\frac{1}{(9-d)} r^a{}_a$. Similarly, given $Z\in W_1^*$ we have 
\begin{equation}
\label{eq:E*-tranfs}
\begin{aligned}
   \zeta' &= - c \zeta + r\cdot \zeta - u\inn a + s\inn \ta  , \\
   u'&= - c u + r\cdot u - \alpha\inn \zeta 
         - s\inn a + t\inn\ta , \\
   s'&= - c s + r\cdot s - \talpha\inn\zeta
         - \alpha\wedge u - t\inn a , \\
   t'&= - c t + r\cdot t
         - j\alpha\wedge s - j\talpha \wedge u . 
\end{aligned}
\end{equation}
Finally the adjoint commutator 
\begin{equation}
   R'' = \BLie{R}{R'} 
\end{equation}
has components
\begin{equation}
\label{eq:adj-tranfs}
\begin{aligned}
   c'' &= \tfrac13 (\alpha \inn a' - \alpha' \inn a) + \tfrac23 (\talpha' \inn \ta - \talpha \inn \ta') , \\
   r'' &= \BLie{r}{r'} + j\alpha \inn ja' - j\alpha' \inn ja - \tfrac13 (\alpha \inn a' - \alpha' \inn a) \id \\
   	& \qquad \qquad + j\talpha' \inn j\ta - j\talpha \inn j\ta' 
	- \tfrac23 (\talpha' \inn \ta - \talpha \inn \ta') \id , \\
   a'' &= r \cdot a' - r' \cdot a + \alpha' \inn \ta - \alpha \inn \ta', \\
   \ta'' &= r \cdot \ta' - r' \cdot \ta - a \wedge a', \\
   \alpha'' &= r \cdot \alpha' - r' \cdot \alpha + \talpha' \inn a - \talpha \inn a' , \\
   \talpha'' &= r \cdot \talpha' - r' \cdot \talpha -\alpha \wedge \alpha'
\end{aligned}
\end{equation}
Here we have $c'' = \tfrac{1}{9-d} r''^a{}_a$, as $R''$ lies in the $\Edd$ sub-algebra.

The $\Edd\times\bbR^+$ Lie group can then be constructed starting with
$\GL(d,\bbR)$ and using the exponentiated action of $a$, $\ta$,
$\alpha$ and $\talpha$. The $\GL(d,\bbR)$ action by an element $m$ is
standard so
\begin{equation}
\begin{aligned}
   (m\cdot v)^a &= m^a{}_b v^b , &&&
   (m\cdot \omega)_{ab} 
       &= (m^{-1})^c{}_a (m^{-1})^d{}_b \omega_{cd} , &&&
   \text{etc}. 
\end{aligned}
\end{equation}
The action of $a$ and $\ta$ form a nilpotent subgroup of nilpotency
class two. One has 
\begin{equation}
\begin{aligned}
   \ee^{a+\ta}V &=
      v + \left(\omega + i_va\right)
      \\ & \qquad
      + \big(
         \sigma + a\wedge\omega + \tfrac{1}{2}a\wedge i_va
         + i_v\ta \big) 
      \\ & \qquad
      + \big( \tau + ja\wedge\sigma - j\ta\wedge\omega
         + \tfrac{1}{2}ja\wedge a \wedge \omega
         \\ & \qquad \qquad 
         + \tfrac{1}{2}ja\wedge i_v\ta
         - \tfrac{1}{2}j\ta\wedge i_va
         + \tfrac{1}{6}ja\wedge a \wedge i_va
         \big),
\end{aligned}
\end{equation}
with no terms higher than cubic in the expansion. The action of $\alpha$
and $\talpha$ form a similar nilpotent subgroup of nilpotency 
class two with 
\begin{equation}
\begin{aligned}
   \ee^{\alpha+\talpha}V &=
      \big( v + \alpha\inn\omega - \talpha\inn\sigma
            + \tfrac{1}{2}\alpha\inn\alpha\inn\sigma
            \\ & \qquad \qquad 
            + \tfrac{1}{2}\alpha\inn\talpha\inn\tau
            + \tfrac{1}{2}\talpha\inn\alpha\inn\tau
            + \tfrac{1}{6}\alpha\inn\alpha\inn\alpha\inn\tau
            \big)
      \\ & \qquad
      + \left( \omega + \alpha\inn\sigma + \talpha\inn\tau
            + \alpha\inn\alpha\inn\sigma \right)
      \\ & \qquad
      + \left( \sigma + \alpha\inn\tau\right)
      + \tau . 
\end{aligned}
\end{equation}
A general element of $\Edd\times\bbR^+$ then has the form 
\begin{equation}
   M\cdot V = \ee^\lambda\,\ee^{\alpha+\talpha}\,\ee^{a+\ta}
       \,m\cdot V , 
\end{equation}
where $\ee^\lambda$ with $\lambda\in\bbR$ is included to give a
general $\bbR^+$ scaling. 


\subsection{Some tensor products}
\label{sec:Eddmaps}

We also define two tensor products. We have the map into the adjoint 
\begin{equation}
   \oadj : W_1^* \otimes W_1 \to W_{\text{ad}} .
\end{equation}
Writing $R=Z\oadj V$ we have 
\begin{equation}
\label{eq:EE*-adj}
\begin{aligned}
	c &= - \tfrac13 u\inn \omega 
		- \tfrac23 s\inn \sigma - t \inn \tau ,\\
	r &= v \otimes \zeta - ju \inn j\omega
		+ \tfrac13 (u\inn\omega) \id 
                - js\inn j\sigma
		+ \tfrac23 (s\inn\sigma) \id
		- jt\inn j\tau ,\\
	\alpha &= v \wedge u +  s\inn\omega + t\inn\sigma ,\\
	\talpha &= - v \wedge s - t\inn\omega , \\
	a &= \zeta \wedge \omega + u\inn \sigma + s\inn \tau ,\\
	\ta &= \zeta \wedge \sigma + u\inn \tau . 
\end{aligned}
\end{equation}

We can also consider the space
$W_2$ as given in table~\ref{tab:gen-tensors}. Taking 
\begin{equation}
\begin{aligned}
   W_2 &= F^* 
       \oplus \Lambda^4F^*
       \oplus (F^*\otimes \Lambda^6F^*)
       \oplus (\Lambda^3F^*\otimes\Lambda^7F^*)
       \oplus (\Lambda^6F^*\otimes\Lambda^7F^*) , \\
   Y &= \lambda + \kappa + \mu + \nu + \pi ,
\end{aligned}
\end{equation}
we have that the symmetric map $W_1 \otimes W_1 \rightarrow W_2$ is
\begin{equation}
\label{eq:EE-N}
\begin{aligned}
\lambda &= v \inn \omega' + v' \inn \omega ,\\
\kappa &= v \inn \sigma' + v' \inn \sigma - \omega \wedge \omega' ,\\
\mu &= \left(j\omega\wedge\sigma' + j\omega'\wedge\sigma\right)
	- \tfrac14 \left(\sigma\wedge\omega' + \sigma'\wedge\omega\right) \\
	&\qquad \qquad + (v \inn j\tau) + (v \inn j\tau')  - \tfrac14 (v\inn \tau' + v' \inn \tau),\\
\nu &= j^3 \omega \wedge \tau' + j^3 \omega' \wedge \tau - j^3 \sigma \wedge \sigma' ,\\  
\pi &= j^6  \sigma \wedge \tau' + j^6 \sigma' \wedge \tau  ,
\end{aligned}
\end{equation}
%


\subsection{$\Hd$ and $O(d)$}
\label{app:Hd}

%

Given a positive definite metric $g$ on $F$, which for
convenience we take to be in standard form $\delta_{ab}$ in frame
indices, we can define a metric on $W_1$ by  
\begin{equation}
   G(V,V) = v^2 + \tfrac{1}{2!}\omega^2 
      + \tfrac{1}{5!}\sigma^2 + \tfrac{1}{7!}\tau^2 ,
\end{equation}
where $v^2=v_av^a$, $\omega^2=\omega_{ab}\omega^{ab}$, etc as
in~\eqref{eq:*norm}. Note that this metric allows us to identify
$W_1\simeq W_1^*$. 

The subgroup of $\Edd\times\bbR^+$ that leaves the metric is invariant
is $\Hd$, the maximal compact subgroup of $\Edd$ (see
table~\ref{tab:coset}). The corresponding Lie algebra is
parametrised by   
\begin{equation}
\label{eq:Hd-alg}
\begin{aligned}
   N &= n + b + \tilde{b} \in
      \Lambda^2F^* \oplus \Lambda^3F^* \oplus \Lambda^6F^*  , 
\end{aligned}
\end{equation}
and embeds in $W_{\text{ad}}$ as 
\begin{equation}
\label{eq:Hd-embed}
\begin{aligned}
   c &= 0 , \\
   r_{ab} &= n_{ab} , \\
   a_{abc} = - \alpha_{abc} &= b_{abc}, \\
   \tilde{a}_{a_1 \dots a_6} = \tilde{\alpha}_{a_1 \dots a_6} 
      &= \tilde{b}_{a_1 \dots a_6} , 
\end{aligned}
\end{equation}
where indices are lowered with the metric $g$. Note that $n_{ab}$
generates the $O(d)\subset\GL(d,\bbR)$ subgroup that preserves
$g$. Concretely a general group element can be written as 
\begin{equation}
\label{eq:Haction}
   H\cdot V = \ee^{\alpha+\talpha}\,\ee^{a+\ta}\,h\cdot V , 
\end{equation}
where $h\in O(d)$ and $a$ and $\alpha$ and $\ta$ and $\talpha$ are
related as in~\eqref{eq:Hd-embed}. 

Finally we note that the double cover $\dHd$ of $\Hd$ has a
realisation in terms of the Clifford algebra
$\Cliff(d;\bbR)$. Consider the gamma matrices $\gamma^a$ satisfying
$\{\gamma^a,\gamma^b\}=2g^{ab}$. The $\Hd$ Lie algebra can be realised
on $\Cliff(d;\bbR)$ spinors by taking
\begin{equation}
\label{eq:Hd-cliff}
   N = \tfrac{1}{2}\left(\tfrac{1}{2!}n_{ab}\gamma^{ab} 
        - \tfrac{1}{3!}b_{abc}\gamma^{abc} 
        - \tfrac{1}{6!}\tilde{b}_{a_1\dots a_6}\gamma^{a_1\dots a_6}
        \right). 
\end{equation}
Again $n_{ab}$ generates the $\Spin(d)$ subgroup of $\dHd$. 


\subsection{Type II $\GL(d-1,\bbR)$ and $O(d-1)$ subgroups of
  $\Edd\times\bbR^+$} 
\label{app:typeII}

We can identify two distinct $\GL(d-1,\bbR)$ subgroups of $\Edd$
appropriate to type IIA and type IIB. 

For type IIA, $\GL(d-1,\bbR)$ is a subgroup of the $\GL(d,\bbR)$ group
used to define the $\Edd\times\bbR^+$ group in
section~\ref{app:Edd-def}. We simply decompose the $d$-dimensional
space as 
\begin{equation}
   F \simeq L \oplus \bbR , 
\end{equation}
with a $\GL(d-1,\bbR)$ action on $L$. Concretely, if we write the
$\GL(d,\bbR)$ index $a=(1,i)$ then the $\GL(d-1,\bbR)$ Lie algebra
$p\in L\otimes L^*$ embeds as 
\begin{equation}
\label{eq:IIA-embed}
   r^i{}_j = p^i{}_j . 
\end{equation}
Under this decomposition one has
\begin{equation}
\begin{aligned}
   W_1 &=  L \oplus L^* \oplus \Lambda^5 L^*
          \oplus \left(L^*\otimes \Lambda^6L^* \right) 
          \oplus \Leven L^*  , \\
   W_{\text{ad}} &= \bbR \oplus \bbR \oplus \left(L\otimes L^*\right) 
          \oplus \Lambda^2L^* \oplus \Lambda^2L^* \\
          & \qquad \quad   \oplus \Lambda^6L^* \oplus \Lambda^6L^*
          \oplus \Lodd L^* \oplus \Lodd L^*,
\end{aligned}
\end{equation}

For type IIB the embedding is slightly more complicated. We decompose $\GL(d,\bbR)$ under a $\GL(d-2, \bbR)\times\SL(2,\bbR)$ subgroup, i.e. we decompose
$F$  as a $d-2$-dimensional space $A$ and $2$-dimensional space $B$. We then identify 
\begin{equation}
   F \simeq A \oplus B , \qquad \hat{L} = A \oplus \Lambda^2B^* ,
\end{equation}
where the $\GL(d-1,\bbR)$ action acts on $\hat{L}$ and under $\SL(2,\bbR)$ we have $\Lambda^2B^*\simeq \bbR$ (this is needed for $\hat{L}$ to form a representation of $\GL(d-1,\bbR)$). Writing indices
$a=(1,2,\hat{\imath})$, the $\GL(d-1,\bbR)$ Lie algebra element
$\hat{p}\in \hat{L} \otimes\hat{L}^*$ embeds as 
\begin{equation}
\label{eq:IIB-embed}
   r^{\hat{\imath}}{}_{\hat{\jmath}} 
       = \hat{p}^{\hat{\imath}}{}_{\hat{\jmath}} , 
   \qquad 
   \alpha^{\hat{\imath}12} = \hat{p}^{\hat{\imath}}{}_1 , 
   \qquad 
   a_{\hat{\imath}12} = \hat{p}^1{}_{\hat{\imath}} ,
   \qquad 
   r^1{}_1 = r^2{}_2 = - \tfrac{1}{2}\hat{p}^1{}_1 . 
\end{equation}
Decomposing under the $\GL(d-2, \bbR)\times\SL(2,\bbR)$ subgroup and then recombining the terms into $\GL(d-1, \bbR)\times\SL(2,\bbR)$ representations we find
\begin{equation}
\begin{aligned}
  W_1  &= \hat{L} \oplus \Lambda^3 \hat{L}^* 
  	\oplus \big( \hat{L}^*\otimes \Lambda^6\hat{L}^* \big) 
	\oplus \big[ B \otimes \big( \hat{L}^* \oplus \Lambda^5\hat{L}^* \big) \big] , \\
  W_{\text{ad}} &= \bbR \oplus \big(B \otimes B^*\big)_0 \oplus \big(\hat{L}\otimes \hat{L}^*\big)
  	\oplus \Lambda^4\hat{L}^* \oplus \Lambda^4\hat{L} \\
          & \qquad \quad \oplus \big[ B \otimes \big(\Lambda^2\hat{L}^* 
          	\oplus \Lambda^2\hat{L} 
		\oplus \Lambda^6\hat{L}^* 
          	\oplus \Lambda^6\hat{L} \big) \big] .
\end{aligned}
\end{equation}
After breaking the $SL(2,\bbR)$ action on $B$ this becomes
\begin{equation}
\begin{aligned}
	W_1 &=  \hat{L} \oplus \hat{L}^* \oplus \Lambda^5\hat{L}^*
          \oplus \big(\hat{L}^*\otimes \Lambda^6\hat{L}^* \big) 
          \oplus \Lodd\hat{L}^*  , \\
	W_{\text{ad}} &= \bbR \oplus \bbR \oplus \big(\hat{L}\otimes \hat{L}^*\big) 
          \oplus \Lambda^2\hat{L}^*  \oplus \Lambda^2\hat{L} \\
          &\qquad \quad \oplus \Lambda^6\hat{L}^* \oplus \Lambda^6\hat{L} 
          \oplus \Leven\hat{L}^* \oplus \Leven\hat{L} . 
\end{aligned}
\end{equation}

The corresponding embeddings of $O(d-1)$ in $\Hd$ follow from the
intersection of the embedding~\eqref{eq:Hd-embed}
with~\eqref{eq:IIA-embed} and~\eqref{eq:IIB-embed}. The $\Hd$ algebra
element decomposes as 
\begin{equation}
\label{eq:Hd-alg-II}
\begin{aligned}
   N &= q + s + \tilde{s} + t^- 
       \in \ \Lambda^2L^* \oplus \Lambda^2L^* \oplus \Lambda^6L^*
          \oplus \Lodd L^* , \\
     &= \hat{q} + \hat{s} + \hat{\tilde{s}} + \hat{t}^+ 
       \in   \Lambda^2\hat{L}^* \oplus \Lambda^2\hat{L}^* 
          \oplus \Lambda^6\hat{L}^* 
          \oplus \Leven \hat{L}^* . 
\end{aligned}
\end{equation}
Lifting to a $\Spin(d-1)$ action, it is important to note that
$\Cliff(d-1;\bbR)$ for the type IIB spinors does not embed in
$\Cliff(d;\bbR)$; only the spin group $\Spin(d-1)$
embeds. Concretely, in both cases, one can decompose the
$\Cliff(d;\bbR)$ spinors under $\gamma^1$ by  
\begin{equation}
   \gamma^1 \epsilon^\pm = \pm \epsilon^\pm . 
\end{equation}
Each spinor $\epsilon^\pm$ then transforms under the $\Spin(d)$ group
generated by 
\begin{equation}
\begin{aligned}
   \hat{\gamma}^{ij} &= \gamma^{ij} & && && \text{type IIA} \\
   \hat{\gamma}^{ij} 
      &= \begin{cases} 
           \gamma^{\hat{\imath}\hat{\jmath}} 
               & \text{if $i=\hat{\imath}, j=\hat{\jmath}$} \\
           \gamma^{\hat{\imath}12} 
               & \text{if $i=\hat{\imath}, j=1$} \\
           - \gamma^{\hat{\jmath}12} 
               & \text{if $i=1, j=\hat{\jmath}$}
          \end{cases}
          & && && \text{type IIB}
\end{aligned}
\end{equation}
One then has the Clifford action for the type IIA decomposition 
\begin{equation}
\begin{aligned}
   N \epsilon^\pm &= \tfrac{1}{2}\Big(\tfrac{1}{2!}q_{ab}\hat{\gamma}^{ab} 
        \mp \tfrac{1}{2!}s_{ab}\hat{\gamma}^{ab} 
        - \tfrac{1}{6!}\tilde{s}_{a_1\dots a_6}\hat{\gamma}^{a_1\dots a_6}
        \Big) \epsilon^\pm
        \\ & \qquad \qquad \qquad 
        - \tfrac{1}{2}\sum_n \tfrac{1}{n!}(\pm)^{[(n+1)/2]}
             t^-_{a_1\dots a_n}\hat{\gamma}^{a_1\dots a_n} \epsilon^\mp     
\end{aligned}
\end{equation}
and
\begin{equation}
\begin{aligned}
   N \epsilon^\pm &= \tfrac{1}{2}\Big(
        \tfrac{1}{2!}\hat{q}_{ab}\hat{\gamma}^{ab} 
        \mp \tfrac{1}{2!}\hat{s}_{ab}\hat{\gamma}^{ab} 
        - \tfrac{1}{6!}\hat{\tilde{s}}_{a_1\dots a_6}
           \hat{\gamma}^{a_1\dots a_6} \Big) \epsilon^\pm
        \\ & \qquad \qquad \qquad 
        - \tfrac{1}{2}\sum_n \tfrac{1}{n!}(\pm)^{[(n+1)/2]}
             \hat{t}^+_{a_1\dots a_n}\hat{\gamma}^{a_1\dots a_n} 
             \epsilon^\mp . 
\end{aligned}
\end{equation}
for type IIB. 



\end{document}